\documentclass[amssymb,amsmath,showpacs,superscriptaddress,preprint]{revtex4-1}
\usepackage{graphicx}
\usepackage{amsmath}
\usepackage{gensymb}
\usepackage{xcolor}
\usepackage{blindtext}
\usepackage[allcolors=blue]{hyperref}%
 
\begin{document}
\title{Stress-tunable abilities of glass forming and mechanical
amorphization}
\author{Xinxin Li$^\#$}
\affiliation{Songshan Lake Materials Laboratory, Dongguan 523808, China}
\affiliation{Department of Mechanical Engineering, The University of Hong Kong, Pokfulam Road, Hong Kong SAR, China}
\author{Baoshuang Shang$^\#$}
\email{shangbaoshuang@sslab.org.cn} 
\affiliation{Songshan Lake Materials Laboratory, Dongguan 523808, China}
\author{Haibo Ke}
\email{kehaibo@sslab.org.cn}
\affiliation{Songshan Lake Materials Laboratory, Dongguan 523808, China}
\author{Zhenduo Wu}
\affiliation{City University of Hong Kong (Dongguan), Dongguan 523808, China}
\author{Yang Lu}
\affiliation{Department of Mechanical Engineering, The University of Hong Kong, Pokfulam Road, Hong Kong SAR, China}
\author{Haiyang Bai}
\email{hybai@iphy.ac.cn}
\affiliation{Songshan Lake Materials Laboratory, Dongguan 523808, China}
\affiliation{Institute of Physics, Chinese Academy of Sciences, Beijing 100190, China}
\affiliation{
Center of Materials Science and Optoelectronics Engineering, University of
Chinese Academy of Sciences, Beijing 100049, China
}
\author{Weihua Wang}
\email{whw@iphy.ac.cn}
\affiliation{Songshan Lake Materials Laboratory, Dongguan 523808, China}
\affiliation{Institute of Physics, Chinese Academy of Sciences, Beijing 100190, China}
\affiliation{
Center of Materials Science and Optoelectronics Engineering, University of
Chinese Academy of Sciences, Beijing 100049, China
}
\date{\today}
\begin{abstract}
Mechanical amorphization, a widely observed phenomenon, has been utilized to synthesize novel phases by inducing disorder through external loading, thereby expanding the realm of glass-forming systems. 
Empirically, it has been plausible that mechanical amorphization ability consistently correlates with glass-forming ability. 
However, through a comprehensive investigation in binary, ternary, and quaternary systems combining neutron diffraction, calorimetric experimental approaches and molecular dynamics simulation, we demonstrate that this impression is only partly true and we reveal that the mechanical amorphization ability can be inversely correlated with the glass forming ability in certain cases To provide insights into these intriguing findings, we present a stress-dependent nucleation theory that offers a coherent explanation for both experimental and simulation results.
Our study identifies the intensity of mechanical work, contributed by external stress, as the key control parameter for mechanical amorphization, rendering the ability to tune this process. 
This discovery not only unravels the underlying correlation between mechanical amorphization and glass-forming ability but also provides a pathway for the design and discovery of new amorphous phases with tailored properties.
\end{abstract}
\maketitle

\section{Introduction}

Mechanical amorphization is a phenomenon frequently encountered in the context of crystalline materials\cite{Li2022Amorphization}, encompassing metallic\cite{Zhao2021Amorphization,Wang2021Deformation}, ionic\cite{Meade1990static}, and covalent compounds\cite{Tang2021synthesis}. This process not only broadens the spectrum of attainable glassy states\cite{McMillan2005density}, but also offers a path to the discovery of entirely new amorphous phases\cite{RosuFinsen2023medium}. 
Typically, mechanical amorphization is driven by external forces, including milling\cite{Koch1983PreparationO,Suryanarayana2001,PhysRevB.56.R11361}, shock waves\cite{Chen2003shock,Zhao2018ShockinducedAI}, ultrasonic vibrations\cite{Li2023ultrafast}, or various forms of deformation\cite{He2016insitu,Zhang2018Amorphous,Luo2019plasticity}. These processes lead to the formation of intricate structures, resulting in materials that exhibit remarkable combinations of mechanical and functional properties.
These properties may encompass enhanced strength\cite{Shang2021ultrahard}, ductility\cite{Luo2019plasticity,Hu2023amorphous}, and impact toughness\cite{Huang2020natural}. Consequently, comprehending and revealing the controlling factors of mechanical amorphization ability (MAA) are of paramount importance.

As an alternative amorphization method, glass forming refers specifically to that produced through fast quenching. Empirically, the MAA of a material has been observed to plausibly correlate with its glass forming ability (GFA)\cite{Sharma2007criterion,Sharma2008EffectON}, that is, the ability of the achieved critical cooling rate when quenching the melt bypassing crystallization. 
In systems with stronger GFA (the slower critical cooling rate required), mechanical amorphization tends to occur more readily, e.g., a large negative heating of mixing and atomic size mismatch. Consequently, glass-forming systems have often been the focus of investigations in mechanical amorphization\cite{Koch1983PreparationO,Zhang2018Amorphous,Hu2023amorphous,Hemley1988PressureinducedAO}.
However, it's crucial to recognize that these two non-equilibrium processes follow distinct pathways to achieve amorphous structures.
In the glass-forming process, amorphous structures are preserved from the liquid state through rapid quenching, effectively circumventing crystallization. 
In contrast, mechanical amorphization introduces disorder into a crystalline structure. The connection between these two processes remains enigmatic\cite{Suryanarayana2018,Zhao2023amorphization}.
Furthermore, the obtained amorphous composition ranges in these two processes are inconsistent. Notably, in the case of mechanical alloying, materials with good MAA usually tend to be at the point of equal atomic ratio due to the relatively low Gibbs free energy of amorphous phase\cite{Lund2004MolecularSO}, which contrasts with the eutectic points typically associated with glass-forming systems\cite{PhysRevLett.91.115505}. 
Understanding the disparity between GFA and MAA, as well as unraveling the mechanisms and control factors governing mechanical amorphization, are pivotal questions that remain unresolved.

In this study, we employed mechanical alloying methods in both experimental and simulation settings to explore a range of glass-forming systems. 
Our primary objective was to delve into the mechanisms that underlie MAA and its correlation with GFA. 
Surprisingly, our experimental results contradicted the conventional wisdom. 
Contrary to empirical impressions, we observed an abnormal correlation pattern between MAA and GFA across all the systems investigated herein. 
In the simulation phase of our research, we further probed the relationship between GFA and MAA. 
Intriguingly, we found that this correlation could be modulated by varying the loading conditions. 
Under low-stress conditions, GFA exhibited a positive correlation with MAA, aligning with the empirical impression. 
However, under high-stress conditions, the correlation reversed, consistent with our experimental findings. 
To shed light on these unexpected results, we introduced a stress-dependent nucleation theory that offers a comprehensive explanation for both experimental and simulation outcomes. 
Our study identified the intensity of mechanical work contributed by external stress as the pivotal control parameter governing mechanical amorphization. 
The newfound tunability offers opportunities to subtly tailor amorphous-crystalline nanostructures with superior performance by mechanical amorphization.
\section{Experimental}

\subsection{Materials}

For ribbon preparation, binary CuZr (with nominal compositions of Zr$_{50}$Cu$_{50}$, Zr$_{55}$Cu$_{45}$, Zr$_{60}$Cu$_{40}$, Zr$_{65}$Cu$_{35}$), ternary CuZrAl (with nominal compositions of Zr$_{45}$Cu$_{45}$Al$_{10}$, 
Zr$_{47.5}$Cu$_{42.5}$Al$_{10}$, Zr$_{50}$Cu$_{40}$Al$_{10}$, Zr$_{52.5}$Cu$_{37.5}$Al$_{10}$, Zr$_{55}$Cu$_{35}$Al$_{10}$,
Zr$_{57.5}$Cu$_{32.5}$Al$_{10}$, Zr$_{60}$Cu$_{30}$Al$_{10}$, Zr$_{62.5}$Cu$_{27.5}$Al$_{10}$, Zr$_{65}$Cu$_{25}$Al$_{10}$, 
Zr$_{67.5}$Cu$_{22.5}$Al$_{10}$, Zr$_{70}$Cu$_{20}$Al$_{10}$, Zr$_{72.5}$Cu$_{17.5}$Al$_{10}$), 
quaternary CuZrNiAl (with nominal compositions of Zr$_{45}$Cu$_{35}$Ni$_{10}$Al$_{10}$, Zr$_{50}$Cu$_{30}$Ni$_{10}$Al$_{10}$, 
Zr$_{55}$Cu$_{25}$Ni$_{10}$Al$_{10}$, Zr$_{60}$Cu$_{20}$Ni$_{10}$Al$_{10}$) alloys ingots were initially produced by arc melting mixtures of the raw metals Cu, Zr, Al, and Ni (with a purity of $\geq$ 99.9 wt. \%) in an Ar atmosphere (with a purity of $\geq$ 99.9999\%) purified with a Ti getter. 
All compositions are given in atomic percent. The ingots were flipped and remelted four times to ensure compositional homogeneity.
Subsequently, glassy ribbons with a thickness of approximately 35 $\mu$m were produced by melt spinning on a single Cu roller at a wheel surface speed of 35 m s$^{-1}$ in a high-purity Ar atmosphere.

For powder preparation, the mixtures of elemental powders were mechanically alloyed. 
First, high purity metal powders ($\geq$ 99.99 at. \%) were weighted according to the corresponding nominal composition and blended in a plastic vessel at a speed of 50 rpm for 24 h under a high-purity Ar atmosphere. 
The blends were then mechanically alloyed in a planetary ball mill at speeds of 450 and 350 rpm (QM-2SP20; apparatus factory of Nanjing University, Nanjing, China) under a protective atmosphere of high-purity Ar atmosphere. 
A WC ball with a diameter of 20 mm and a ball to powder weight ratio (BPR) of 10:1 was used. 
As a parallel comparison, the milling experiments of speed 450 rpm and BPR 7:1 was performed. 
Approximately 5 g of the as-milled powders were taken out from the WC vials in intervals of 5, 8, 11, 14, 17, 20, 23, 26, 30, 35, 40, 45, 50, 55, 60, 65, 70, and 75 h for phase evolution and thermal analysis. 
Generally, it is accepted that only the presence of a diffraction halo indicates the full amorphization of element powders during ball milling. 
From a calorimetric point of view, amorphous phase exhibits an exothermic behavior at the elevated temperatures, called, crystallization (Fig. \ref{fig:S1} middle part).
Due to the accumulation of disordered structures (amorphous phase) during ball milling, the value of crystallization enthalpy $H_{x}$ increased monotonically with milling times and reached a maximum one $H_x^\text{max}$ until full amorphization 
(Fig.\ref{fig:S1} lower part). 
Furthermore, the process of mechanical amorphization can be characterized by the reduced parameter $H_x/H_x^\text{max}$, which is consistent with the evolution of XRD patterns (Fig.\ref{fig:S1} upper part). 
Furthermore, high-resolution TEM images of as-milled amorphous phases presented the similar maze-like patterns with that of as-cast counterpart (Fig.\ref{fig:S2}). 
Hence, the MAA can be evaluated by amorphization time $t_a$\cite{Ge2017}, corresponding to the moment that $H_x/H_x^\text{max}$ is equal to 1. The related details of the remaining alloys were seen from Figs. \ref{fig:S3},\ref{fig:S4},\ref{fig:S5},\ref{fig:S6}.

\subsection{Thermal analysis}

Each tested sample of approximately 25 mg was analyzed using a synchronous thermal analyzer in alumina crucibles under a high-purity of Ar atmosphere (STA-449 F3, NETZSCH, Germany). 
To ensure the reliability of the data, temperature and enthalpy were calibrated with an indium and a zinc standard specimen, giving an accuracy of $\pm$0.1 K and $\pm$0.01 mW, respectively. 
The scanning procedure was conducted at a heating rate of 20 K min$^{-1}$ until complete melting, and then cooled to ambient temperature at 40 K min$^{-1}$. 
A second run using the same procedure was used as a baseline for subtraction from the first run. The glass transition temperature ($T_g$), onset of crystallization temperature ($T_x$), and liquidus temperature ($T_l$) were determined as the intersection points of the tangents at the inflection points. 
The enthalpy of crystallization ($H_x$) was measured from the area between two curves. 
To investigate the kinetics of crystallization, the scanning procedure was carried out up to 823 K with various heating rates of 5, 10, 20, 50, 100, and 200 K min$^{-1}$ (DSC 8000, PE, USA) to determine the $T_x$ dependent on heating rates. 
A second run using the same procedure was used as a baseline for subtraction from the first run.
Heat capacity ($C_p$) of alloys considered here was obtained by comparing with that of a sapphire standard sample. 
Identical measurement procedures were performed on the empty pan as a baseline to be subtracted from the sample and the sapphire. The specific heat capacity of the sample can be determined by
\begin{equation}
C_p^{\text{sample}}=C_p^{\text{Sapphire}}\times \frac{M_\text{sapphire}}{M_\text{sample}}
\times \frac{Q_\text{sample}-Q_\text{pan}}{Q_\text{Sapphire}-Q_\text{Pan}}
\end{equation}
where $M_i$ and $Q_i$ are the mass and heat flow of sample, empty pan, and sapphire respectively; and $C_p^\text{Sapphire}$ represents the heat capacity of the standard sapphire.

\subsection{High-resolution X-ray diffraction}

Samples were placed in a borosilicate capillary with an outer diameter of 0.5 mm and a wall thickness of 0.01 mm. 
High-resolution X-ray diffraction (HRXRD) measurements were performed with a rotating Ag anode at a wavelength  $\lambda$ of 0.056 nm and a beam size of 8 mm $\times$ 0.4 mm at a power of 9 kW (Smart Lab, Rigaku, Japan). 
The sample scanning speed is 1.5$\degree$ per minute and the scanning range  is from 20$\degree$ to 90$\degree$. 
Before amorphous samples were tested, a borosilicate capillary is firstly examined by HRXRD as a baseline.

\subsection{Neutron diffraction}

Neutron diffraction measurements were performed on the Multiple Physics Instrument (MPI) at China Spallation Neutron Source (CSNS)\cite{Xu2021multiphyiscs}. 
Approximately 3 g of samples were placed in a vanadium can with an inner diameter of 8.9 mm and thickness of 0.25 mm. Single Diffraction pattern was measured for 6 h at ambient condition in the high-flux mode. The scattering data were analyzed using the Mantid software, which corrected, normalized and collated the information from the 7 banks\cite{Arnold2014Mantiddata}. To obtain a high $Q$ resolution over a large distance range, the instrument resolution was initially assessed using a Si powder (SRM 640f, NIST) before amorphous samples were tested. The $dQ/Q$ was expected to reach 0.39\% at 1 \AA. The normalized shape profiles of single pixel for the Si (422) peak show that the resolution of FWHM is 0.36\%.

\subsection{Electron microscopy}

TEM specimens were carefully ion milled with 3 keV Ar ions for about 5 h at the liquid nitrogen temperature (PIPS II-695.c, Gatan, USA). High-resolution transmission electron microscopy (HRTEM) measurements were performed using a field emission gun TEM (JEM F200, JEOL, Japan) at 200 kV.

\subsection{Molecular dynamic simulation}

\subsubsection{Initial sample preparation and loading process}

The mechanical amorphization in various glass-forming systems was investigated through molecular dynamics simulations.
Initially, polycrystalline structures for the Cu-Zr system was generated using the Atomsk package\cite{Hirel2015Atomsk}. 
The simulation box dimensions were set to 10 nm $\times$ 10 nm $\times$ 10 nm, and eight grains of the corresponding alloy phase were created using Voronoi tessellation. The average grain diameter was about 5 nm. For the Cu-Zr system, two different alloy phases, namely CuZr B2 phase and CuZr$_2$ phase, were used in the polycrystal structure. The CuZr sample consisted of approximately 58000 atoms, and the CuZr$_2$ sample contained around 52000 atoms. Interatomic interactions were modeled using semi-empirical potentials based on the embedded atom model (EAM) for the Cu-Zr system\cite{Mendelev2019development}. Molecular Dynamics simulations were conducted using the open-source software LAMMPS\cite{Thompson2021LAMMPSA}. The simulation time step was set to 2 fs, and periodic boundary conditions were applied in all dimensions.
The OVITO package\cite{Stukowski2009visualization} was employed for atomic visualization. To enhance statistical significance and estimate error bars, five independent samples with random crystal orientations were used for each system.
The initial sample was first minimized at 0 K to balance the grain boundary and then maintained at 300 K with the isothermal–isobaric (NPT) ensemble at ambient pressure using the Nos\'e-Hoover thermostat and barostat\cite{Nos1984AUF,Martyna1994ConstantPM} for 200 ps. 
The ball milling process can be mimicked by cyclic loading process\cite{Lund2004MolecularSO,Rogachev2022MechanicalAI}, in this study, we employed pure shear oscillatory deformation to investigate the mechanical amorphization phenomenon introducing by mechanical alloying. 
The sinusoidal loading was applied along the z-direction, and opposite loading was applied along the x and y directions. The strain along the z-direction followed the form $\epsilon_{zz}=\gamma_A \sin (2 \pi t/t_p )$, where  represented the strain amplitude, and  denoted the periodic time, which was kept constant at 100 ps (see Fig. \ref{fig:S7}). Additionally, the strain along the x and y directions followed the form $\epsilon_{xx},\epsilon_{yy}=-\gamma_A/2 \sin (2 \pi t/t_p)$. During the loading, the temperature was maintained at 300 K by Nos\'e-Hoover thermostat\cite{Nos1984AUF,Martyna1994ConstantPM}.

\subsubsection{Characteristic of the degree of disorder}

In our study, we assessed the degree of disorder during the loading process by calculating the ``Order". The order bond between two neighboring atoms, denoted by $i$ and $j$, was established using the scalar product:
\begin{equation}
   S_6(i,j) \equiv \frac{\sum^6_{m=-6}{q_{6m}(i) \cdot q_{6m}^*(j)}}{\sqrt{\sum^6_{m=-6}{q_{6m}(i) \cdot q_{6m}^*(i)}}\sqrt{\sum^6_{m=-6}{q_{6m}(j) \cdot q_{6m}^*(j)}}}
\end{equation}
where $q_{6m}$ represents the standard bond-orientations parameter\cite{PhysRevB.28.784}, and $q_{6m}^{*}$ is the corresponding complex conjugate.

To identify an order bond between neighboring atoms $i$ and $j$, we considered $S_6(i,j)>0.7$. 
The local degree of disorder for atom $i$ was determined by a summation of all the order bonds involving atom $i$:
\begin{equation}
   \text{disorder} \equiv 1-\frac{1}{N_c} \sum_{j \in N_c(i)} \Theta(S_6(i,j)-0.7)
\end{equation}
where $\Theta(x)$ is the step function, and $N_c(i)$ represents the number of neighbors of atom $i$. By averaging the ``disorder'' over all atoms in the simulation box, we obtained the ``Disorder''. In a perfect order state, $\text{Disorder}=0$, whereas in a disordered state, Disorder is high and approaches one\cite{Auer2004numerical,Russo2012microscopic}.

\subsection{ Gibbs free energy difference between the liquid and crystalline states and interfacial energy }

The calculated difference in the Gibbs free energy $\Delta g$ between the liquid and crystalline states is given by
\begin{equation}
   \Delta g  \equiv   \Delta  H -T\Delta S 
             =    (\Delta H_f-\int_{T}^{T_f} \Delta C_p^{l-c}(T)dT)-
            T(\Delta S_f - \int_T^{T_f} \frac{\Delta C_p^{l-c}(T)}{T}dT)
\end{equation}

where $\Delta H_f$ is the heat of fusion, $\Delta S_f$ is the entropy of fusion, rendering $\Delta S_f=\Delta H_f/T_f$. $T_f$, the temperature at which the Gibbs free energy of the liquid and the crystalline states are equal, was taken to be the temperature at which the endothermic peak is maximum during melting. The heat capacity of a crystal well above the Debye temperature can be described by\cite{Glade2000ThermodynamicsOC}
\begin{equation}
   C_p^c(T)=3R+aT+bT^2 
\end{equation}
The heat capacity of an undercooled liquid can be described by\cite{Gallino2010KineticAT}[41]
\begin{equation}
   C_p^l(T)=3R+cT+dT^{-2}
\end{equation}
where $R$ is gas constant, rendering R= 8.3145 J g atom$^{-1}$ K$^{-1}$, and $a$, $b$, $c$,
and $d$ are fitting constants. The constants for both fits to the specific heat capacity data for each of alloy systems were summarized in Supplementary Table \ref{tab:s2}. 
The difference in the specific heat capacity of the liquid and the
crystalline states, $\Delta C_p^{l-c}$, was shown in Fig. \ref{fig:S27}. 
The counterparts for binary and quaternary were depicted in Figs. \ref{fig:S16}a and \ref{fig:S17}a. Correspondingly, Figs. \ref{fig:S16}b, \ref{fig:S17}b, and \ref{fig:S28} show the difference in the
Gibbs free energy $\Delta {g}$ between the liquid and crystalline states for binary, quaternary, and ternary alloy systems respectively.
In general, interfacial energy $\gamma_{c/a}$ is difficult to be experimentally measured. 
Turnbull pointed that a relation exists between the liquid-crystal
interfacial energy and the heat of fusion\cite{Turnbull1950FormationOC}.
Such a relation can be described by
\begin{equation}
   \gamma_{c/a}=K \Delta H_f V^{-2/3}N_A^{-1/3} 
\end{equation}
where $V$ is the gram-atomic volume of the crystalline phase, $N_A$ is the
Avogadro constant, $K$ is the fitting parameter. Turnbull also found that
while data for most of metals fit on a line with slope $K=0.45$, Ge, Sb, and
Bi were best fit by a line with slope 0.32\cite{Turnbull1950FormationOC}. Hence, $K=0.45$ is used in this work.

\subsection{Determination of total work \texorpdfstring{$W$}{Lg}}

Regardless of shear stress\cite{He2016insitu,Hu2023amorphous} or hydrostatic stress\cite{Hemley1988PressureinducedAO,Zhao2015pressure}, each of stress state can drive amorphization. Both stress components (including the maximum shear stress  or hydrostatic stress P) are determined by the generalized Hooke’s law, as detailed in Supplementary Note 1.
Patel and Cohen\cite{Patel1953CriterionFT} were the first to quantitatively study the effect of stress states on the phase transformation under quasi-static loading by rationalizing the total work, $W= P \epsilon_v+\tau \gamma$.
Here, $\epsilon_v$ is volume shrinkage between amorphous and crystalline phase. The atomic volume $v_a$ of amorphous phase is determined by the relationship proposed by Ma et al.\cite{Ma2009PowerlawSA} based on experimental values of $q_1$ (the position of first peak in plot of $S(Q)$ versus of Q, see Figs. \ref{fig:S8}a and \ref{fig:S9}). The  of crystalline phase is estimated from the equilibrium phase diagram using the lever rule. $\gamma$ is the shear strain, rendering  (G is the shear modulus\cite{Wang2012TheEP}). Hence, $W=P\frac{v_a-v_c}{v_c}+\frac{\tau^2_{max}}{G}$.

\section{Results}

\subsection{ GFA of binary, ternary, and quaternary glass-forming systems}
\begin{figure}
    \centering
    \includegraphics[width=0.8\textwidth]{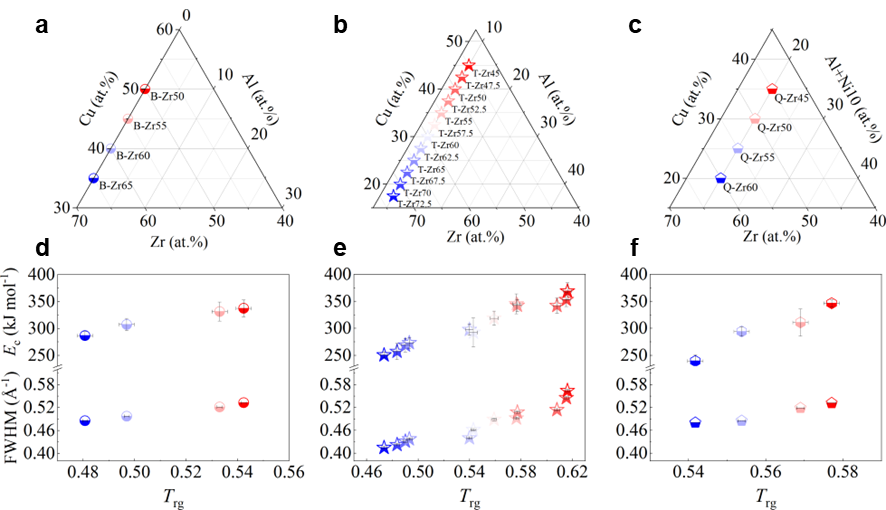}
    \caption{Composition contours \textbf{(a-c)} and GFA indicators \textbf{(d-f)} for binary ZrCu \textbf{(a, d)}, ternary ZrCuAl \textbf{(b, e)}, and quaternary ZrCuNiAl \textbf{(c, f)} alloy systems with various Zr contents. 
    FWHM is calculated from the full width at half maximum of the first peak of $S(Q)$. $T_{rg}$ and $E_c$ denote the reduced temperature and crystallization activation energy calculated by $T_g/T_l$ and Kissinger equation respectively. 
    The Kissinger plots of alloy systems considered herein with various heating rates are detailed in the Supplementary Figure \ref{fig:S13}. All the related values can be seen from the Supplementary Table \ref{tab:S1}.}
    \label{fig:1}
\end{figure}

In this study, we selected a series of Zr-based alloys as representative materials. These alloys are renowned for their superior GFA and have been extensively examined in the prior studies\cite{PhysRevLett.94.205501,PhysRevLett.115.165501,Li2008matching,Zhu2016,PhysRevLett.106.125504,Luo2018UltrastableMG,Yu2013ultrastable}. 
Specifically, we considered binary ZrCu, ternary ZrCuAl, and quaternary ZrCuNiAl alloys with varying Zr content, denoted as B-Zr$_x$, T-Zr$_x$, and Q-Zr$_x$ ($x$ representing the Zr atomic content), respectively. The composition variations within the binary ZrCu, ternary ZrCuAl, and quaternary ZrCuNiAl alloys are detailed in Figs. \ref{fig:1} a, b, and c respectively. 

We commenced by characterizing the GFA of each composition using three well-established parameters: the calorimetric parameter reduced temperature $T_{rg}$ ($=T_g/T_l$)\cite{Turnbull1969UnderWC}, the energetic parameter crystallization activation energy $E_c$\cite{Wang2012TheEP}, and the structural parameter full width at half maximum (FWHM) of the first peak in $S(Q)$\cite{Li2021datadriven}.
Detailed experimental procedures are outlined in the Experimental section and the Supplementary Materials (Supplementary Note 2 and Fig. \ref{fig:S8}). For alloys with stronger GFA, all $T_{rg}$, $E_c$, and FWHM are expected to be larger. In our investigated systems, these parameters exhibit a consistent correlation with each other, as depicted in Figs.\ref{fig:1} d, e, and f for binary, ternary, and quaternary alloys respectively. 
Specifically, the FWHM of T-Zr$_{72.5}$ and T-Zr$_{45}$ are a minimum and maximum value of 0.41504 and 0.56429 \AA, respectively, which are among the reported range of $0.28 \sim 0.67$ \AA   for as-deposited ZrCuAl glass library\cite{Li2021datadriven}. 
Obviously, the values of T$_{rg}$ and  for the bulk glass former T-Zr$_{45}$ are 0.616 and 0.408, larger than those of the marginal glass former T-Zr$_{72.5}$, respectively. From a structural and calorimetric point of view, these results indicate that GFA of ternary alloy system herein decreases with the increasing Zr content. 
Similarly, a relationship between GFA and Zr content of alloy materials also exists in binary and quaternary systems as shown in Figs.\ref{fig:1} d and f respectively. The related details of binary and quaternary alloys are seen from the Fig. \ref{fig:S9}.

\subsection{Reversal of MAA and tunable correlation between MAA and GFA for glass-forming systems}

In our investigation of mechanical amorphization, we employed a technique known as mechanical alloying (MA). 
This process involves subjecting a mixture of powders to high-energy collisions by milling balls within a confined vessel. 
The continuous input of energy through mechanical milling imparts disorder to the powder materials, facilitating their transformation towards an amorphous phase\cite{Suryanarayana2019MechanicalAA}. 
There are several advantages to using MA for studying mechanical amorphization. 
Firstly, it offers a high degree of control compared to other methods such as shock waves or intense plastic deformation\cite{Ge2017}. Secondly, it enables the attainment of a fully amorphous state in the material\cite{Eckert1997mechanical}. 
Moreover, perhaps the most crucial advantage is that MA provides an effective means to characterize mechanical amorphization ability. Specifically, the time taken for milling to achieve a fully amorphous state can serve as a quantifiable measure of this ability. 
\begin{figure}
    \centering
    \includegraphics[width=0.8\textwidth]{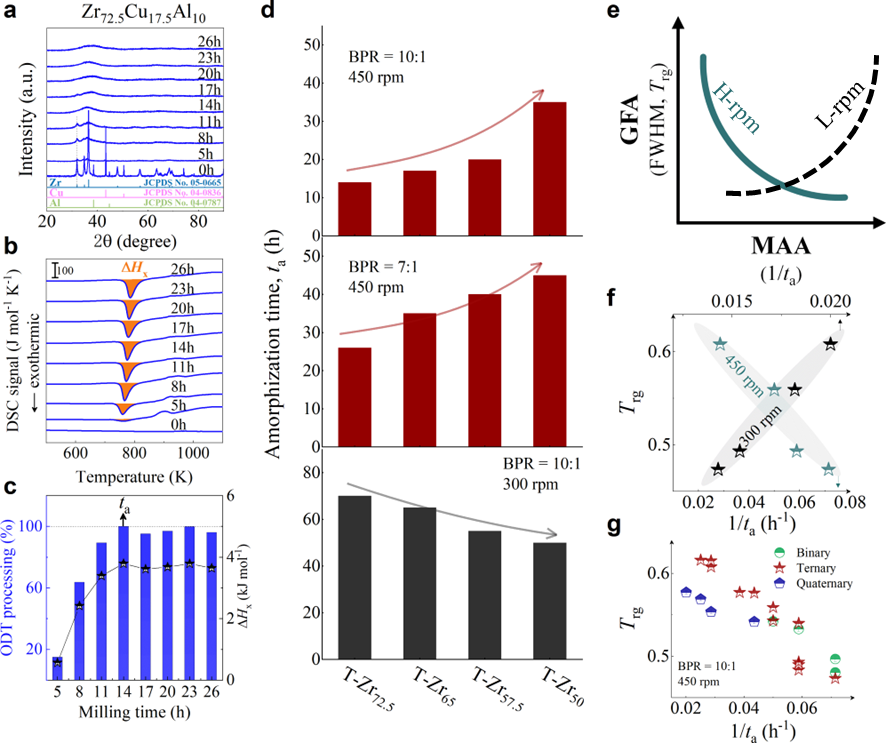}
    \caption{\textbf{(a-c)} Composition contours \textbf{(a)}, GFA indicators \textbf{(b)}, and MAA indicators \textbf{(c)} for ternary ZrCuAl alloy systems with various Zr contents. 
    FWHM is calculated from the full width at half maximum of the first peak of $S(Q)$. $T_{rg}$ and $E_c$ denote the reduced temperature and crystallization activation energy calculated by $T_g/T_l$ and Kissinger equation respectively. 
    The Kissinger plots of alloy systems considered herein with various heating rates are detailed in the Supplementary Fig. \ref{fig:S13}. 
    All the related values can be seen from the Supplementary Table \ref{tab:S1}. $t_a$ is the amorphization time. 
    \textbf{(d, e)} Anomalous correlations exist in plots of FWHM versus $1/t_a$ \textbf{(d)} and $T_{rg}$ versus $1/t_a$ \textbf{(e)} for all alloy systems considered here. 
    The unexpected correlation also exists in plots of $E_c$ versus $E_\text{input}$ (Supplementary Fig. \ref{fig:S15}).
    \textbf{(f)} Illustration of anomalous correlation between GFA indicators ($T_{rg}$ and FWHM) and MAA indicators ($1/t_a$). }
    \label{fig:2}
\end{figure}
As shown in Fig. \ref{fig:2}, XRD patterns (Fig. \ref{fig:2}a) and heating flows (Fig. \ref{fig:2}b) of T-Zr$_{72.5}$ as a function of milling times depict the evolution of amorphization. 
Apparently, all the originals consist of peaks of the elementary metals Cu, Zr, and Al (Fig. \ref{fig:2}a).
With the increasing milling time, these diffraction peaks gradually attenuated and broadened, which is attributed to the introduction of disorder structures with collision-induced deformation. 
Such an interfacial reaction in a metastable state is promoted by continuous cold welding and fracturing of powder particles to create alternating layers with fresh interfaces, and by the generation of the abundant defects during ball milling\cite{Lund2004MolecularSO,Hellstern1986amorphization,PhysRevLett.65.2019}. 
The composition varies negligibly with the advance of milling and keeps nearly with the equimolar ratio between Cu and Zr elements (see Fig. \ref{fig:S22}). 
In addition, no detectable contamination is found during MA. 
From a calorimetric point of view, due to the accumulation of disordered structures during  the ball milling, this metastable phase exhibits an exothermic behavior at the elevated temperatures, called, crystallization (Fig. \ref{fig:2}b). 
Furthermore, as shown in Fig. \ref{fig:2}c, the value of crystallization enthalpy $H_c$ increases monotonically with milling times and reaches a maximum when the whole amorphous phase is formed. 
Remarkably, the amorphous microstructure obtained after ball milling may be different from that obtained by melt-spun. 
Mechanical amorphization via ball milling is indeed a thermodynamic-controlled stochastic process and $t_a$ is statistical index of the global process, similar to incubation time\cite{Turnbull1969UnderWC} during the crystal nucleation process in the undercooled liquid. 
Hence, this difference does not affect that $t_a$, the time required to achieve amorphous phase via ball milling, is capable to discuss the composition dependence of MAA.

Next, we characterized the MAA using the milling time $t_a$ required to
achieve a fully amorphous state Generally, for superior MAA, the $t_a$ is shorter, and vice versa. Another parameter used to characterize MAA is the input energy $E_\text{input}$ for achieving full amorphization during milling, which exhibits consistent results with $t_a$ (see Supplementary Note 3). For a specific component, the MAA, that is, the time of amorphization does change with the milling speed and BPR (Fig. \ref{fig:2}d). 
Intrinsically, the variation of MAA at different milling conditions is ascribed to the underlying mechanical work.
Previous studies have shown that under the same milling experimental
conditions, the difference in MAA of different compositions is mainly
explained by the criterion of GFA\cite{Sharma2008EffectON,Sharma2007criterion,Zou2013effect,Yang2015effect}, in other words, the system with large GFA tends to have a large MAA (Fig. \ref{fig:S26}c).
Phenomenally, the empirical criteria of MAA proposed by Suryanarayana
(Fig. \ref{fig:S26}a) are analogous to the Ionue’s three rules (Fig. \ref{fig:S26}b), such as atomic size difference and negative mixing enthalpy. However, Fig. \ref{fig:2}e show that two types of correlations between MAA and GFA were both found and the positive one shifts toward the negative with the elevated milling speed (Fig. \ref{fig:2}f), underlying the effect of mechanical work. 
Both the GFA indicators, $T_{rg}$ (Fig. \ref{fig:2}g) and FWHM (Fig. \ref{fig:S29}) parameters, exhibit an inverse relationship with MAA indicator, $1/t_a$. 
The similar relationship exists in plots of $E_c$ with $E_\text{input}$ (Fig. \ref{fig:S15}). 
These suggest that, for materials with excellent GFA, the milling time $t_a$ tends to be longer, indicating a poorer MAA (green line of Fig. \ref{fig:2}e). 
This observation challenges the prevailing empirical impression that a higher GFA in an alloy result in a faster mechanical amorphization However, our experimental findings reveal an inverse correlation between MAA and GFA.

\subsection{Crossover between MAA and GFA in molecular dynamic simulation}
\begin{figure}
    \centering
    \includegraphics[width=0.8\textwidth]{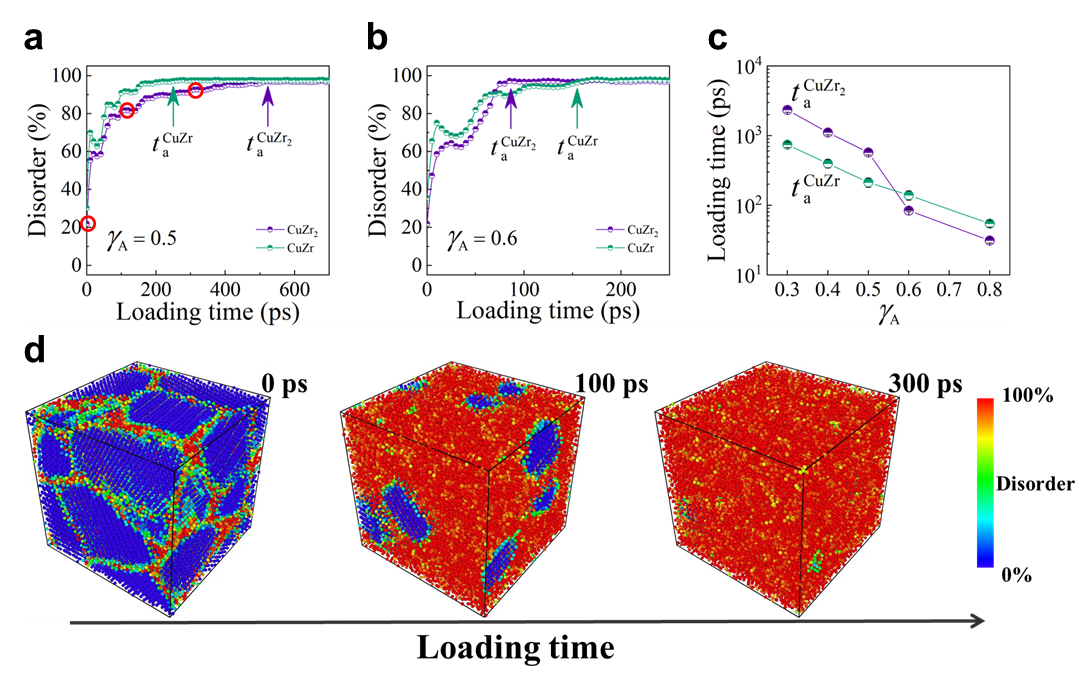}
    \caption{
    \textbf{(a, b)} The Disorder degree as a function of loading time with different loading amplitudes \textbf{(a)} $\gamma_A=0.5$ and \textbf{(b)} $\gamma_A=0.6$ , respectively, for various systems. $t_a^\text{CuZr}$ and $t_a^{\text{CuZr}_2}$ denote the amorphization time for CuZr and CuZr$_2$ systems, respectively. 
    \textbf{(c)} Amorphization time versus amplitude strain $\gamma_A$  for CuZr and CuZr$_2$ systems.
   \textbf{(d)} Evolution of atomic disorder for different loading times, which are illustrated in \textbf{(a)} by red circles. 
    Error bars is comparable with point size in \textbf{(c)}.}
    \label{fig:3}
\end{figure}
In our molecular simulations, we focused on two model alloys, equimolar CuZr and CuZr$_2$, to investigate the phenomenon of mechanical amorphization. These two alloys are well-suited for this study due to their distinct GFA and their extensive examination in molecular dynamics (MD) simulations\cite{Hu2020physical}. The polycrystalline structure comprises approximately 80\% of these two phases.
Notably, CuZr exhibits superior GFA compared to CuZr$_2$. 
And we ensured the construction of polycrystalline structures for both model systems by employing similar grain sizes and crystalline orientations and the corresponding grain boundary energy is comparable (for a comprehensive description of the modeling procedures, please refer to the Experimental section).
Fig. \ref{fig:3}a illustrates the evolution of the Disorder degree over loading time. When the Disorder degree reaches a plateau value, we define this time as the mechanical amorphization time, denoted as $t_a$. 
Consequently, the MAA of a system can be effectively characterized by its corresponding $t_a$. 
In the regular cases, where $t_a$ is shorter, the MAA is considered better. 
For amplitude strain $\gamma_A = 0.5$, we observed that $ t_a^\text{CuZr} < t_a^{\text{CuZr}_2} $, aligning with the conventional impression that strong glass-forming systems tend to exhibit better MAA. 
However, as depicted in Fig.\ref{fig:3}b, for $ \gamma_A = 0.6$, a different narrative unfolds. 
Here, the situation is reversed, with  $ t_a^\text{CuZr} < t_a^{\text{CuZr}_2} $ , consistent with the aforementioned experimental finding that stronger glass-forming systems paradoxically possess poorer MAA. 
This intriguing reversal indicates that the relationship between MAA and GFA is highly sensitive to external loading conditions. 
To delve deeper into this sensitivity, we conducted a series of systematic tests under various loading conditions, as elucidated in Fig.\ref{fig:3}c. These results unequivocally emphasize that the correlation between MAA and GFA is indeed contingent upon the subtle external loading conditions imposed on the system. 
Notably, for small stress, the MAA appears consistent with GFA, whereas for larger stress, the MAA inversely correlates with GFA. Furthermore, Fig.\ref{fig:3}d provides valuable insights into the spatial evolution of the Disorder degree throughout the mechanical amorphization process, hinting at a behavior reminiscent of melting.

\section{Discussion}
\subsection{The characterization of glass forming ability}
\begin{figure}
    \centering
    \includegraphics[width=0.8\textwidth]{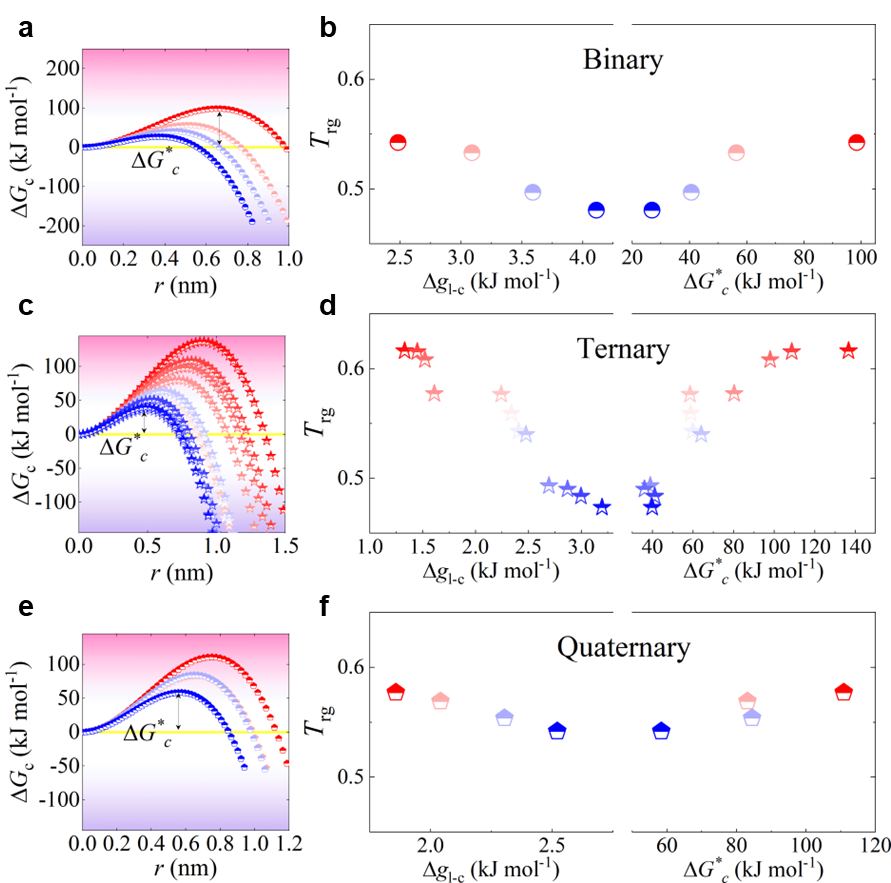}
    \caption{
    Energy penalty associated with nucleation of a crystal embryo \textbf{(a, c, e)} and correlation between GFA indicator $T_{rg}$
    and 
    $\Delta g_{l-c}$ and
    $\Delta G_c^*$
    \textbf{(b, d, f)} for binary \textbf{(a, b)}, ternary \textbf{(c, d)}, and quaternary \textbf{(e, f)} alloy systems, respectively.}
    \label{fig:4}
\end{figure}
The glass forming ability of an alloy system is theoretically attributed to two factors: the resistance to nucleation of crystallization and the resistance to growth of crystallization \cite{Turnbull1969UnderWC,Johnson2016quantifying}.
The resistance to nucleation can be measured by the critical nucleation barrier ($\Delta G_c^* $). Based on classical nucleation theory \cite{Becker1935KinetischeBD}, the nucleation barrier can be expressed as follows:

where $\delta g_{l-c}$ represents the Gibbs free energy difference between the liquid and crystalline states, and $\gamma_{l-c}$ denotes the interfacial energy between the amorphous and crystalline phases. Both  $\gamma_{l-c}$ and $\delta g_{l-c}$ can be determined experimentally (see Methods). As illustrated in Fig. \ref{fig:4}a, c, and e, $\Delta G_c$ evolves with the crystal nucleus size r in binary, ternary, and quaternary systems, respectively. The highest energy barrier $\Delta G^*_c$ in Fig. \ref{fig:4}, is referred to as the critical nucleation barrier. For our investigated systems, we found that all the GFA parameters used in the experiment are positively correlated with the critical nucleation barrier and negatively correlated with the Gibbs free energy difference $\delta \gamma_{l-c}$. This indicates that the GFA of our investigated systems is primarily dominated by the resistance to nucleation.

\subsection{Mechanism of anomalous correlation between MAA and GFA}
\begin{figure}
    \centering
    \includegraphics[width=0.8\textwidth]{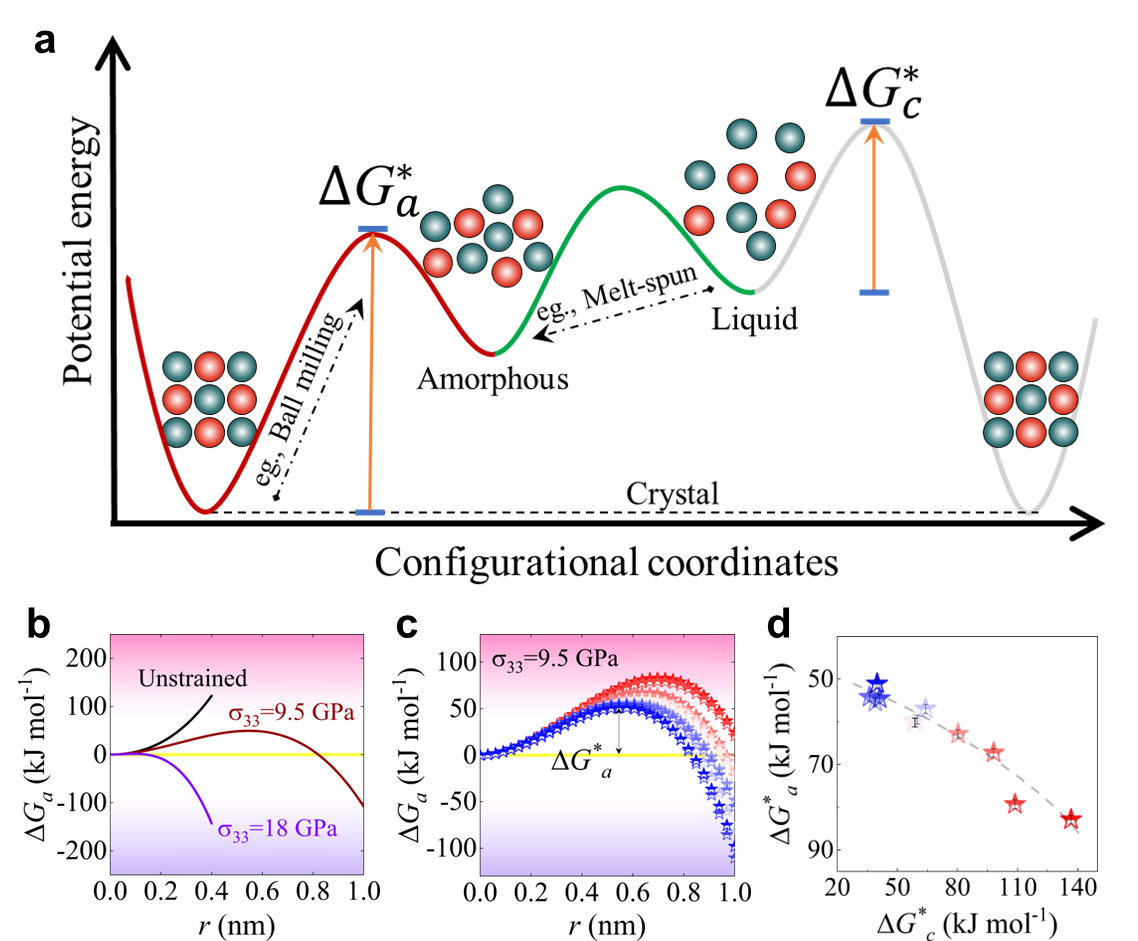}
    \caption{
    \textbf{(a)} Illustrations of two amorphization pathways including mechanical amorphization (labelled by red line) and fast quenching (labelled by green line), and the ability of the latter can be evaluated by suppressing crystallization (labelled by gray line). 
    \textbf{(b, c)} Energy penalty associated with nucleation of an amorphous embryo for ternary ZrCuAl alloy systems, respectively. 
    The impact during ball milling renders it possible to overcome the energy barrier of mechanical amorphization in
    \textbf{(b)}. 
    \textbf{(d)} Correlation of $\Delta G_a^*$ with $\Delta G_c^*$. The dash line indicates a trend that a good as-quenched glassy former with large $\Delta G_c^*$ means superior resistance against crystallization during cooling, 
    which presents a large $\Delta G_a^*$ for hindering the advance of mechanical amorphization.
    The related details of the remaining binary and quaternary alloys systems are seen from the Supplementary figure \ref{fig:S18}. 
    Error bars in \textbf{(d)} mean the standard deviation of data.
}
    \label{fig:5}
\end{figure}
The control factors and mechanisms for modulating MAA remain enigmatic. Notably, the stress state appears to play a pivotal role in reconciling the disparities between experimental and simulation outcomes. From the perspective of the potential energy landscape (see Fig. \ref{fig:5} a), it becomes evident that mechanical amorphization and glass forming represent two fundamentally distinct pathways toward the formation of an amorphous phase. 
One pathway involves the introduction of disorder structures from a crystalline state, while the other revolves around the preservation of disorder structures and the avoidance of crystallization. Preservation of disorder structures in the liquid state can be likened, thermodynamically, to the energy barrier that resists crystallization ($\Delta G_c$),  
In the context of mechanical amorphization, the process of crystalline structure dissolution bears a resemblance to amorphous embryo nucleation phenomena (Fig. \ref{fig:3}d). At a temperature below melting temperature, the amorphization transformation must overcome a high energy barrier, making it impossible to occur under ambient condition. 
However, the high magnitude of the coupled hydrostatic pressure and associated deviatoric component dramatically lowers the energy barrier, rendering the amorphization transformation possible under shock compression. 
Following the Patel and Cohen methodology, the energetics of solid state amorphization of covalently bonded solids, such as silicon\cite{Zhao2016AmorphizationAN,PhysRevB.96.054118}, germanium\cite{Zhao2018ShockinducedAI},
SiC\cite{Zhao2021directional,Levitas2012HighdensityAP}, and B$_4$C\cite{Zhao2021directional}, were analyzed by calculating of the external work varied with pressure and shear. 
In this regard, we introduce a modified version of the classical nucleation theory, widely employed to characterize the transition from a crystalline to an amorphous state\cite{Glade2000ThermodynamicsOC,Zhao2016AmorphizationAN}. 
Within this framework, the energy barrier resisting amorphization, denoted as $\Delta G_a$, can be expressed as:
\begin{equation}
    \Delta G_a = 4 \pi r^2 \gamma_{c/a} +\frac{4}{3} \pi r^3 \Delta g_{c-a}-\frac{4}{3} \pi r^3 W
    \label{eqn:9}
\end{equation}
where  $\Delta g_{c-a}$ signifies the Gibbs free energy difference between the amorphous solid and crystalline states, while $\gamma_{c/a}$  characterizes the interfacial energy between these two phases. $W$ represents the mechanical work induced by external loading. 
Note that, the introduction of crystal defects into metal powders during mechanical amorphization may lead to an increase in the free energy of the crystal phase, surpassing that of the hypothetical amorphous phase. Hence, the third term on the right-hand side of Eq. \ref{eqn:9} represents the contributed energy penalty to overcome the energy barrier of amorphization, including the grain boundaries energy contributed by the exterior stress.
As a result, the most significant energy barrier, denoted as, stands as a pivotal metric for the assessment of a material’s MAA.
Notably, a higher  implies a longer duration required to achieve amorphization, thus signifying a poorer MAA. 
At lower temperatures, particularly below the glass transition temperature,  and  exhibit insensitivity to temperature fluctuations\cite{Turnbull1950FormationOC,Busch1995ThermodynamicsAK}.
We therefore adopt the approximations $\gamma_{c/a} \approx \gamma_{l/c}$ and $\Delta g_{c-a} \approx \Delta g_{l-c}$. 
The formula for mechanical work $W$ depends on the stress conditions. 
Here, we assume the material undergoes uniaxial compressive stress, denoted as $\sigma_{33}$. 
It’s important to note that different stress conditions can alter the value of $W$, but they don’t change the qualitative conclusions. The relationship between $W$ and $\sigma_{33}$ is detailed in the Experimental section.

Fig. \ref{fig:5}c clearly demonstrates that various loading stresses notably influence MAA. Under a fixed loading stress condition, such as $ \sigma_{33} = 9.5$ GPa, different materials exhibit varying MAA levels (see Fig. \ref{fig:5}d), consistent with the experimental results presented in Fig. \ref{fig:2} d. Moreover, the correlation between  and  reaffirms the inverse relationship between MAA and GFA in ternary ZrCuAl alloy systems (Fig. \ref{fig:5}e), in line with the experimental observations (Figs. \ref{fig:2} f and \ref{fig:2} g). 

\subsection{Tunable ability of mechanical amorphization dependent on external stress}
\begin{figure}
    \centering
    \includegraphics[width=0.8\textwidth]{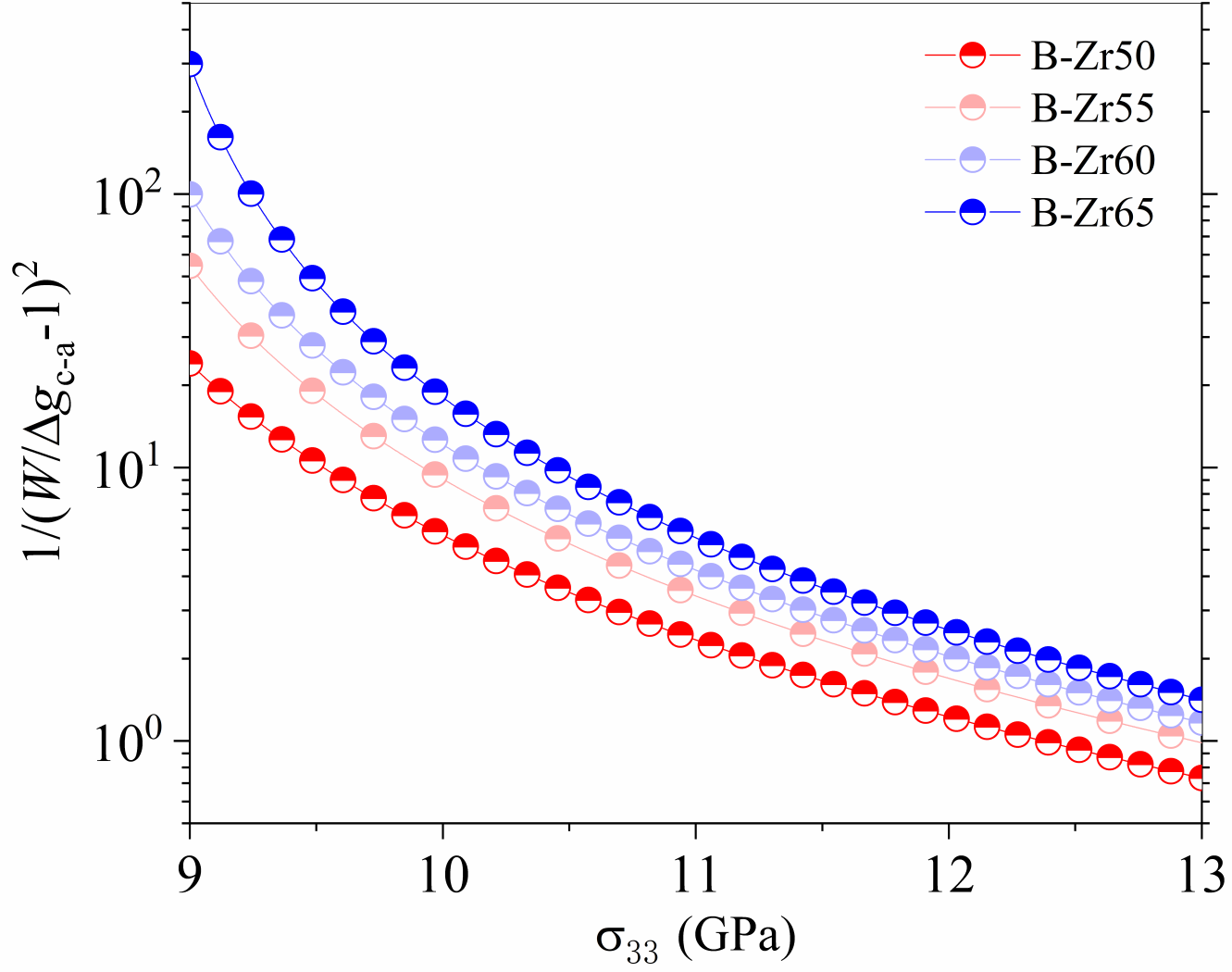}
    \caption{
    The plot of the term 
    $1/(W/\Delta g_{c-a} - 1)^2$
    versus $\sigma_{33}$ 
    for binary alloy system.
}
    \label{fig:6}
\end{figure}
Therefore, using the above calculation frameworks, we can establish the relationship between MAA and GFA as follows:
\begin{equation}
\Delta G_a^* \approx \Delta G_c^* \frac{1}{(W/\Delta g_{c-a}-1)^2}    
\label{eqn:10}
\end{equation}
Evidently, MAA is influenced by both external mechanical work and the composition properties, specifically the energy barrier $\Delta G_c^*$ and the free energy difference $\Delta g_{c-a}$. 
Eq. \ref{eqn:10} can be dissected into two components concerning composition sensitivity.
Firstly, $\Delta G_c^*$ characterizes the GFA of the composition which is insensitive with external loading. 
Secondly, the term  $1/(W/\Delta g_{c-a}-1)$ delineates the interplay between external loading conditions. 
At low stress levels, the external work is comparable with the free energy difference  $\Delta g_{c-a}$ and the term $1/(W/\Delta g_{c-a}-1)$ exhibits sensitive to composition, particularly a positive correlation with $\Delta g_{c-a}$. 
For stronger glass-forming systems, this value tends to be smaller, as exemplified by the notable difference between the marginal glass-former B-Zr$_{65}$ and the good glass-former B-Zr$_{50}$ (about one order, see Fig. \ref{fig:6}).
In contrast, at high stress levels, where $\Delta \ll \Delta g_{c-a}$, the composition insensitivity becomes apparent, leading to significantly reduced differences between B-Zr$_{50}$ and B-Zr$_{65}$. 
In this regime, all values fall within a comparable range of $0.8 \sim 1.1$ (Fig. \ref{fig:6}). 
Therefore, the composition sensitivity of MAA is predominantly governed by $\Delta g_{c-a}$ at low stress levels and shifts towards being dominated by $\Delta G_c^*$ at high stress levels.

\begin{figure}
    \centering
    \includegraphics[width=0.8\textwidth]{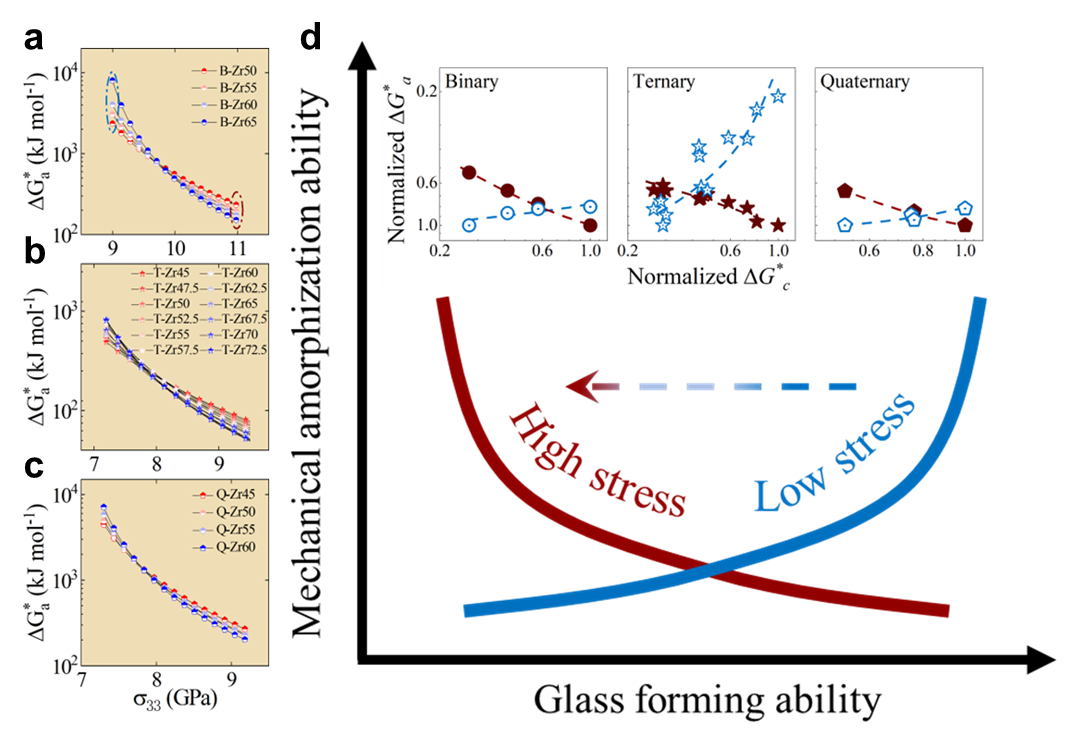}
    \caption{ \textbf{(a-c)} $\Delta G_a^*$ dependence of external stress $\sigma_{33}$ for \textbf{(a)} binary, \textbf{(b)} ternary, and \textbf{(c)} quaternary glass-forming systems considered here. \textbf{(d)} Illustration of tunable correlations between two types of abilities of mechanical amorphization and glass forming with external stress. 
    The high and low stress level are corresponding to the ones circled by red and blue dash line in \textbf{(a)} respectively.
    The inset in \textbf{(d)} presented tunable correlations occurs in each of alloy systems considered here. 
    In terms of glass forming ability, $\Delta G_c^*$ is normalized by the related values of the best glassy former B-Zr$_{50}$, T-Zr$_{45}$, and Q-Zr$_{40}$ for binary, ternary, and quaternary alloy
}
    \label{fig:7}
\end{figure}

We tested a series of loading stresses using Eq. \ref{eqn:10}, as shown in Fig. \ref{fig:7} a-c, all the experimental systems, including binary, ternary, and quaternary alloy systems, reveal a crossover correlation between MAA and GFA. Under low-stress conditions, MAA exhibits a positive correlation with GFA (Fig. \ref{fig:S19}), while under high-stress conditions, MAA shows an inverse correlation with GFA.
This finding aligns with the simulation results (Fig. \ref{fig:3}c) and helps reconcile the discrepancy between current experimental results and established impressions\cite{Sharma2008EffectON,Sharma2007criterion,Zou2013effect}.
Note that, the critical stress level responsible for this reversion varies with the alloy systems. At this stress level, it becomes evident that the MAA becomes entirely independent of GFA, highlighting the sensitivity of the correlation between MAA and GFA to external stress levels (Figs. \ref{fig:7}a-c).
It is intriguing to observe the inverse correlation between MAA and GFA under large stress conditions (Fig. \ref{fig:7}d). This suggests that for materials with poor GFA, mechanical amorphization can be highly efficient, thus offering a novel avenue to expand the range of attainable glassy states \cite{PhysRevB.56.R11361,Eckert1997mechanical}. This finding is particularly fascinating in the context of monatomic glasses\cite{Zhong2014FormationOM,ZHAO2021114018}, where mechanical amorphization, through processing techniques like shocking or various deformation loadings, could potentially offer significant advantages over rapid cooling\cite{Zhao2018ShockinducedAI,He2016insitu,Zhao2016AmorphizationAN}.
Furthermore, this discovery opens up possibilities for generating new amorphous phases, especially in materials with poor GFA, such as amorphous ice\cite{RosuFinsen2023medium}. Additionally, the tunable nature of mechanical amorphization provides valuable insights into controlling amorphization in material design. For instance, the creation of gradient amorphous-crystalline nanostructures with superior performance becomes feasible by mechanical amorphization techniques, such as ultrasonic vibration\cite{Fan2023rapid}.
\section{Conclusion}
In summary, we explored the MAA of materials and its relationship with GFA. 
Surprisingly, we found that the correlation between MAA and GFA is highly sensitive to external loading conditions, reversing under high stress. 
This emphasizes the role of external stress as a key control parameter for MAA. 
Our modified classical nucleation theory revealed that MAA depends on external mechanical work and GFA. 
Notably, under large stress, poor GFA materials could efficiently undergo mechanical amorphization, expanding the range of attainable glassy states and offering exciting possibilities for material design.
\section{Acknowledgements}
  We thank neutron diffraction measurements on the Multiple Physics Instrument (MPI) at China Spallation Neutron Source (CSNS) and the milling equipment supports at Dongguan University of Technology.
 \section{Funding}
  This work was financially supported by Guangdong Major Project of Basic and Applied Basic Research, China (Grant Nos. 2019B030302010), Guangdong Basic and Applied Basic Research Foundation China (Grant Nos.2021A1515111107, 2021B1515140005, and 2022A1515011439), the National Natural Science Foundation of China (Grant Nos.52201176, 52071222, and 52130108), Pearl River Talent Recruitment Program (Grant No.2021QN02C04), Young Talent Support Project of Guangzhou Association for Science and Technology (Grant No.QT2024-041).
\section{Author contributions}
X.X.L. and B.S.S. contributed equally to this work. H.B.K., H.Y.B. and W.H.W. conceived and supervised this work; X.X.L. designed and conducted all the experiments with assistance from Z.D.W.; B.S.S. designed and conducted all the simulations with assistance from X.X.L.; Y.L. assisted in the literature review and theoretical analysis. X.X.L., B.S.S., H.B.K., and H.Y.B. wrote the paper with input and comments from all authors. 
\bibliographystyle{unsrt} 

 \pagebreak
 
 \section*{Supplementary}
 \renewcommand{\thefigure}{S\arabic{figure}}
 \renewcommand{\theequation}{S\arabic{equation}}
 \renewcommand{\thesection}{S\arabic{section}}
 \renewcommand{\thetable}{S\arabic{table}}
 \setcounter{figure}{0}
 \setcounter{equation}{0}
 \setcounter{section}{0}
 \setcounter{table}{0}
 \section{Supplementary Notes}
 
 \subsection{Supplementary Note 1: shear stress and hydrostatic pressure}
 According to the generalized Hook’s law, $\sigma_{ij} = C_{ij33} \epsilon_{33} $ , where $C_{ijkl}$
presents the elastic constants, $\epsilon_{33}$ presents the uniaxial strain.
Correspondingly, the hydrostatic pressure $P$ and maximum shear stress $\tau_\text{max}$
can be determined by $P = \frac{1+\nu}{3(1-\nu)}\sigma_{33}$
 and $\tau_\text{max} = \frac{1-2\nu}{2(1-\nu)}\sigma_{33}$, where $\sigma_{33}$
is the impact stress. 
In terms of ZrCu-based metallic glass systems, the Poisson’s ration $\nu$ is under the range of $0.3 \sim 0.32$.  Hence, the value of v is 0.305 for all of alloy systems herein.

\subsection{Supplementary Note 2: Determination of crystallization activation
energy by Kissinger equation}

\begin{figure}[!]
    \centering
    \includegraphics[width=0.7\textwidth]{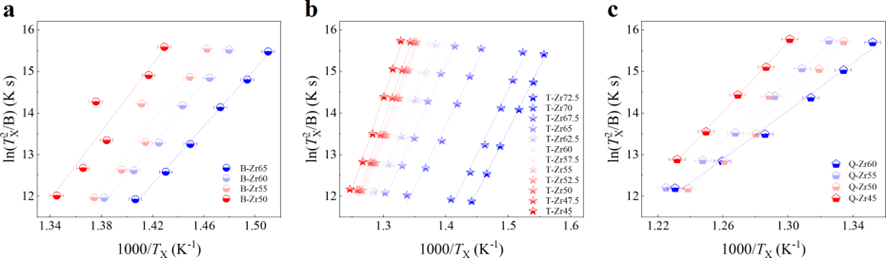}
    \caption{
     Kissinger plots of binary \textbf{(a)}, ternary \textbf{(b)}, and quaternary \textbf{(c)} alloy systems with various heating rates to determine crystallization activated energy.
    }
    \label{fig:S13}
\end{figure}

The apparent activation energy $E_c$ of crystallization is determined under
isochronal heating conditions using the following Kissinger equation
\begin{equation}
    \ln T^2/B=\frac{E_c}{k_B T}+C
    \label{eqn:S1}
\end{equation}
where $k_B$ is the Boltzmann’s constant, B is the heating rate (K s$^{-1}$), $C$ is the
constant and T is the specific temperature dependence of heating rate. 
Generally, T is denoted as the temperature corresponding to the beginning or to the peak of the exothermic crystallization event detected by DSC.
Then, through plotting $\ln(T^2/B)$ as a function of $1/T$, the quantity $E_c$ can be determined by the slope. In this work, $T$ is taken as the onset crystallization temperature $T_x$ and $E_c$ is understood as crystallization activation energy in
this case. 
Correspondingly, Fig. \ref{fig:S13} shows the valid of this dependence.

\subsection{Supplementary Note 3: Estimation of the input energy $E_\text{input}$ in a planetary milling system }

\begin{figure}[!]
    \centering
    \includegraphics[width=0.7\textwidth]{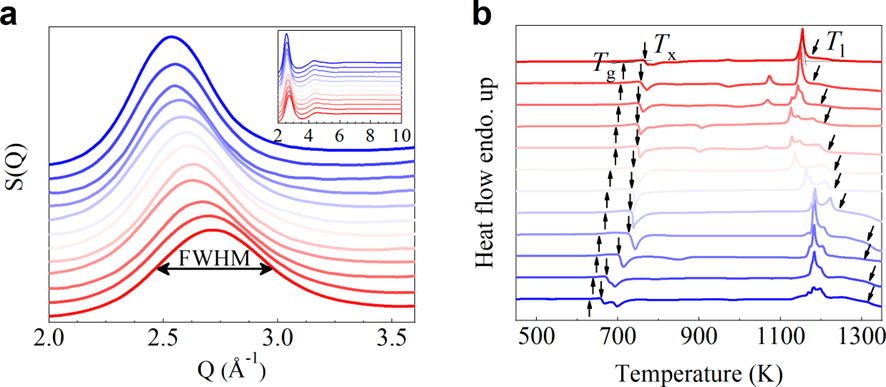}
    \caption{
     \textbf{a} $S(Q)$ obtained from neutron diffraction and \textbf{b} DSC traces at a heating rate of 20 K min$^{-1}$ for ternary ZrCuAl alloy systems. FWHM is calculated from the full width at half maximum of the first peak of $S(Q)$ in a. $T_g$ , $T_x$ and $T_l$ denote transition point of glass transition, crystallization and melting, which are determined by the onset of the transformation as defined by the intersection of the black lines in \textbf{b}.
    }
    \label{fig:S8}
\end{figure}

\begin{figure}[!]
    \centering
    \includegraphics[width=0.7\textwidth]{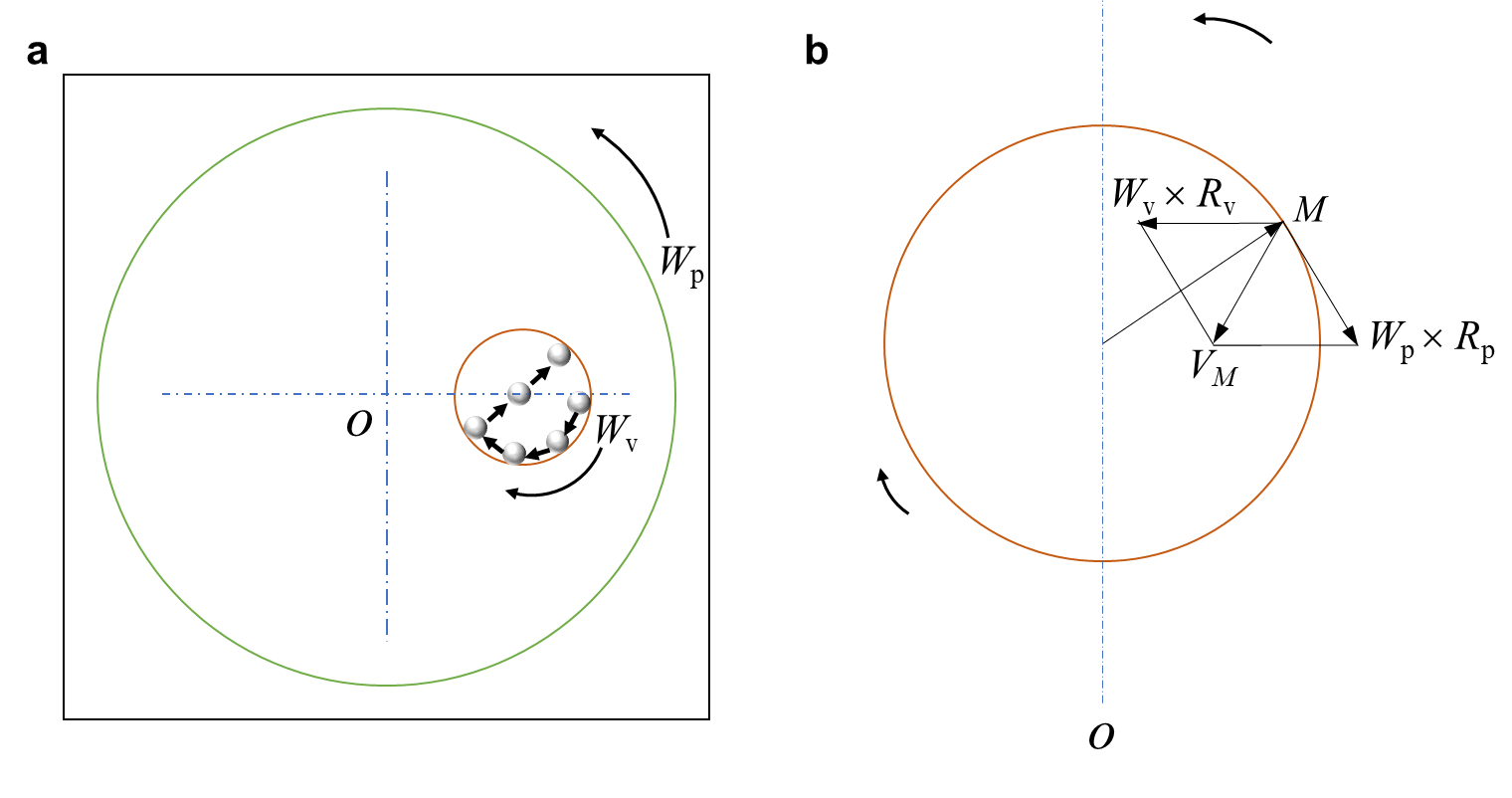}
    \caption{
    \textbf{a)} Geometry scheme of planetary ball milling systems and  \textbf{b)} absolute velocity $V_M$ of one peripheral point M from the top perspective. $W_P$ and $W_V$ , and RP and RV present the absolute angular velocity and radius of the milling plate and steel vial respectively.
    }
    \label{fig:S20}
\end{figure}

In terms of a planetary milling, the geometry of one vial can be schematized in Fig. \ref{fig:S20}. It can be expressed that the absolute velocity of one peripheral point $M$ can be depicted by the following equation:

\begin{equation}
    V_M=[(W_P R_P)^2+(W_V R_V)^2+ 2 W_P R_P W_V R_V \cos{\theta}]^{1/2}
    \label{eqn:S2}
\end{equation}

where $W_P$ and $W_V$ (RPM, rotations per minute) present the absolute angular velocity of the milling plate and steel vial respectively; $R_P$ and $R_V$ (mm) present vectorial distances from the center of the plate to the center of the vial and radius of the vial respectively; $\theta$ is the angle formed by the vectors of $\vec{R}_P$  and $\vec{R}_V$.
It is assumed that the steel ball moves without sliding at the point $M$ (Fig. \ref{fig:S20}). At a specific moment, it is launched towards the opposite inner wall. And the ball moves back to the wall after a short hit and is accelerated by the vial at the next launch cycle. Note that the previous hypotheses are realistically starting from the moment at which a thin powder layer covered the ball. Hence, the condition for the ball detached from the inner wall of vial is expressed by:
\begin{equation}
   \cos{\theta_b} =\frac{W_V^2 R_V}{W_P^2 R_P}
   \label{eqn:S3}
\end{equation}
Combined Eqs. \ref{eqn:S2} and \ref{eqn:S3}, the absolute velocity of one ball leaving the wall can be expressed by:
\begin{equation}
   V_b=[(W_P R_P)^2+W_V^2(R_V-d_b/2)^2+2(1-2W_V/W_P)]^{1/2} 
   \label{eqn:S4}
\end{equation}
where $d_b$ (mm) refers to the diameter of one ball. When attached solidly
again with the wall, the velocity of the ball after a hit, $V_s$ , equal to that of
the inner wall and can be given by:
\begin{equation}
V_s = [(W_P R_P )^2 + W_V^2 (R_V - d_b /2)^2 + 2W_P R_P W_V (R_V - d_b /2)]^{1/2}
\label{eqn:S5}
\end{equation}
When the ball is launched, the energy can be shown as:
\begin{equation}
   E_b=\frac{1}{2} m_b V_b^2
   \label{eqn:S6}
\end{equation}
where $m_b$ (g) refers to the mass weight of one ball. After collision, a fraction of energy is released in ways of deformation of powder materials and instantaneous temperature rise of systems. The remaining energy of
the ball becomes:
\begin{equation}
E_s = \frac{1}{2} m_b V_s^2.
\label{eqn:S7}
\end{equation}
Hence, the released energy $\Delta E \equiv =E_b-E_s$ of one ball during a series of collision events
can be expressed by:
\begin{equation}
\Delta E =-m_b [W_V^3(R_V-d_b/2)/W_P+W_P R_P W_V] (R_V-d_b/2)   
\label{eqn:S8}
\end{equation}
In general, a number of balls, $N_b$ , widely used to accelerate the process of amorphization, hinder each other so that Eq. \ref{eqn:S8} must be considered the effect of filling degree of vial by introducing a related empirical parameter, $\phi_b$ . 
To simplification, two boundary cases is: i) $\phi_b=1$ for only one or few balls and ii) $\phi_b =0$ when the vial is completely filled by the balls, no movement occurs. 
Correspondingly, Eq. \ref{fig:S8} is modified as in a realistic experiment:
\begin{equation}
\Delta E^* = \phi_b \Delta E
\label{eqn:S9}
\end{equation}
where $\Delta E^*$ refers to the kinetic energy of one ball in the system including $N_b$ balls. Hence, the total released energy $P$ can be given by:
\begin{equation}
P=\Delta E^* \phi_b N_b f_b
\label{eqn:S10}
\end{equation}
where $f_b$ refers to the frequency of launching for ball, which is proportions the relative rotation speed, $f_b = K(W_P - W_V )/2\pi$. Therefore, Eq. \ref{eqn:S10} is modified as normalizing by powder weight $m_p$ and
multiplying by milling time t denoted as $E_\text{input} \equiv P_t/Km_P$:
\begin{equation}
   E_\text{input} =-\phi_b N_b m_b t(W_P-W_V)[\frac{W_V^3(R_V-d_b/2)}{W_P}+W_P R_P W_V](R_V-d_b/2)/2\pi m_p
   \label{eqn:S11}
\end{equation}
Note that, the above model can be used to determine the released energy by collision between balls and vial walls, while the focused energy transferred to powder materials is apparently only a portion of the derived released energy $E_\text{input}$. 
In spite of the restrictions and hypotheses, the aforementioned model was used to rationalize the previously reported experimental results.

\subsection{Supplementary Note 4: Deduction for $\phi_b$ expression}

As far as $\phi_b$ is concerned, it is found convenient to express it as a function
of two parameters $n_v$ and $n_s$: 
\begin{equation}
   n_v=N_b/N_{b,v} 
   \label{eqn:S12}
\end{equation}
Where $N_{b,v}$ refers to the number of balls that can be contained in a simple
cubic arrangement in the vial and is given by the following estimation
\begin{equation}
N_{b,v}= \pi D_v^2 H_v /4d_b^3 
\label{eqn:S13}
\end{equation}
Where $H_v$ and $D_v$ is the height and diameter of the vial;
\begin{equation}
   n_s=N_b/N_{b,s} 
   \label{eqn:S14}
\end{equation}
Where $N_{b,s}$ refers to the number of balls needed to cover one third of the
inner surface in a simple cubic arrangement and is shown as
\begin{equation}
   N_{b,s} = \pi (D_v-d_b)H_v/3d_b^2
   \label{eqn:S15}
\end{equation}
In order to derive a simple analytical expression for $\phi_b$ , it is assumed that i) $\phi_b = 1$ for $n_v = 0$ (vial is completely empty);
ii) $\phi_b = 0$ for $n_v = 1$ (vial is completely filled up); iii) $\phi_b$ is close to 1 (e.g. 0.95) for $n_s = 1$ (this assumed that the reciprocal hindering of the ball is negligible until one third of the inner surface wall is not covered).
According to above assumptions, a simple formulation of $\phi_b$ can be estimated by
\begin{equation}
   \phi_b=1-n_v^\epsilon 
\end{equation}
Where $\epsilon$ refers to a parameter dependent of ball diameter that can be
evaluated by the assumption (iii):
\begin{equation}
   0.95=1-(N_{b,s}/N_{b,v})^\epsilon 
\end{equation}

\subsection{Supplementary Note 5: mechanical amorphization type}

\begin{figure}[!]
    \centering
    \includegraphics[width=0.7\textwidth]{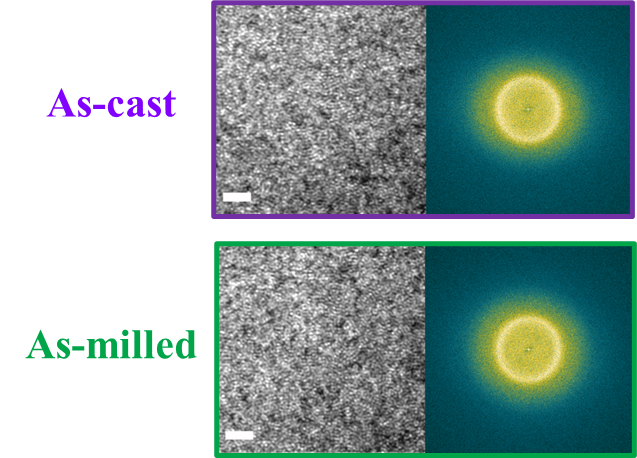}
    \caption{
    HRTEM images and selected area electron diffraction (SEAD) patterns of as-cast (purple line) and as-milled (green line) Zr$_{45}$Cu$_{45}$Al$_{10}$ glassy former.  The scale bar is 2 nm.
    }
    \label{fig:S2}
\end{figure}

\begin{figure}[!]
    \centering
    \includegraphics[width=0.7\textwidth]{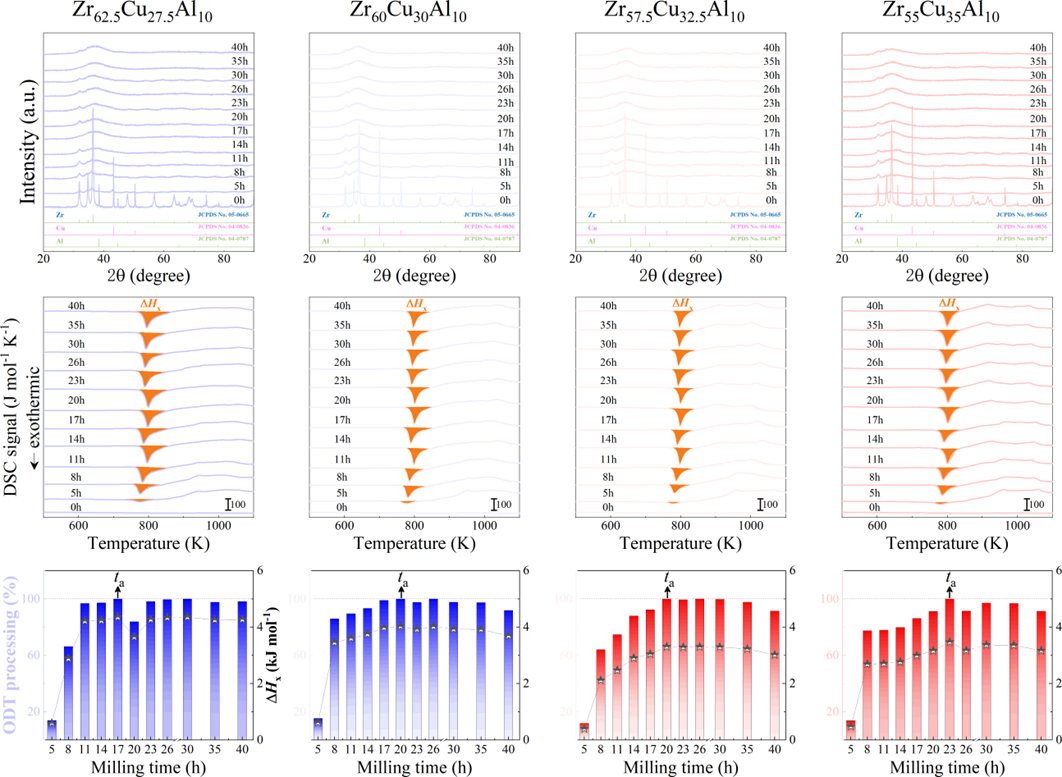}
    \caption{
     XRD pattens (upper part), DSC traces (middle part) at a heating rate of 20 K min$^{-1}$, and processing (lower part) for mechanical amorphization of the T-Zr$_{62.5}$, T-Zr$_{60}$, T-Zr$_{57.5}$, and T-Zr$_{55}$ alloys via high-energy ball milling of pure element powders.
    }
    \label{fig:S3}
\end{figure}

\begin{figure}[!]
    \centering
    \includegraphics[width=0.7\textwidth]{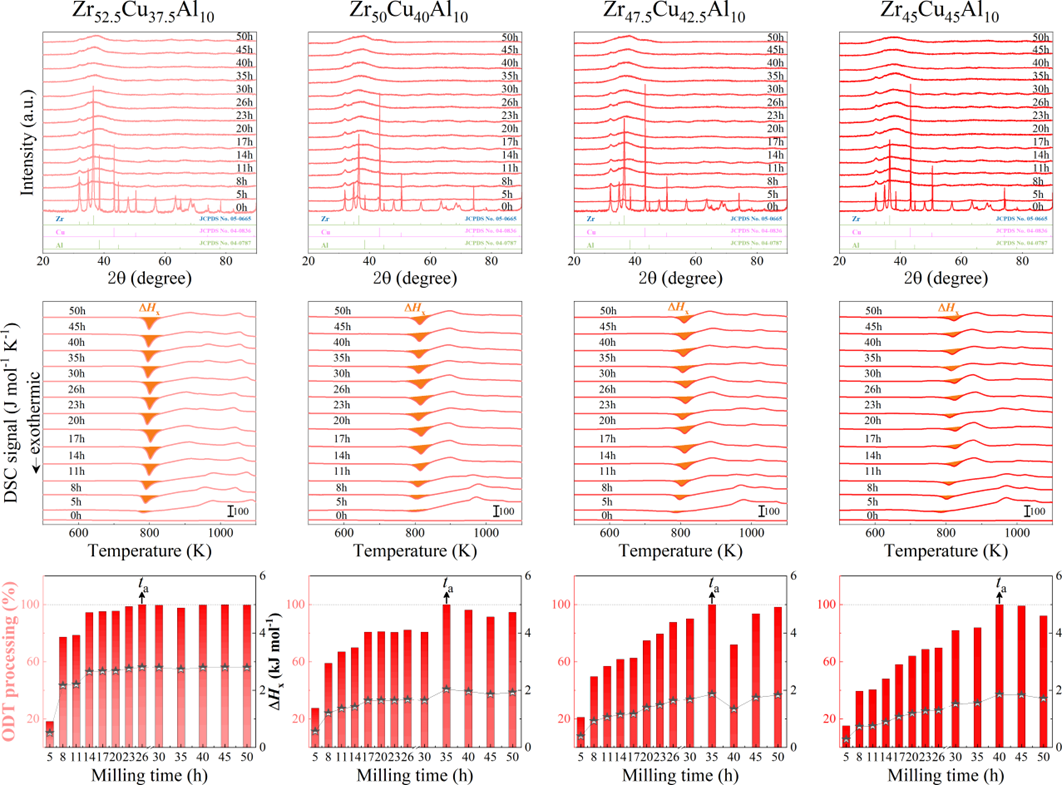}
    \caption{
     XRD pattens (upper part), DSC traces (middle part) at a heating rate of 20 K min$^{-1}$, and processing (lower part) for mechanical amorphization of the T-Zr$_{52.5}$, T-Zr$_{50}$, T-Zr$_{47.5}$, and T-Zr$_{45}$ alloys via high-energy ball milling of pure element powders.
    }
    \label{fig:S4}
\end{figure}

\begin{figure}[!]
    \centering
    \includegraphics[width=0.7\textwidth]{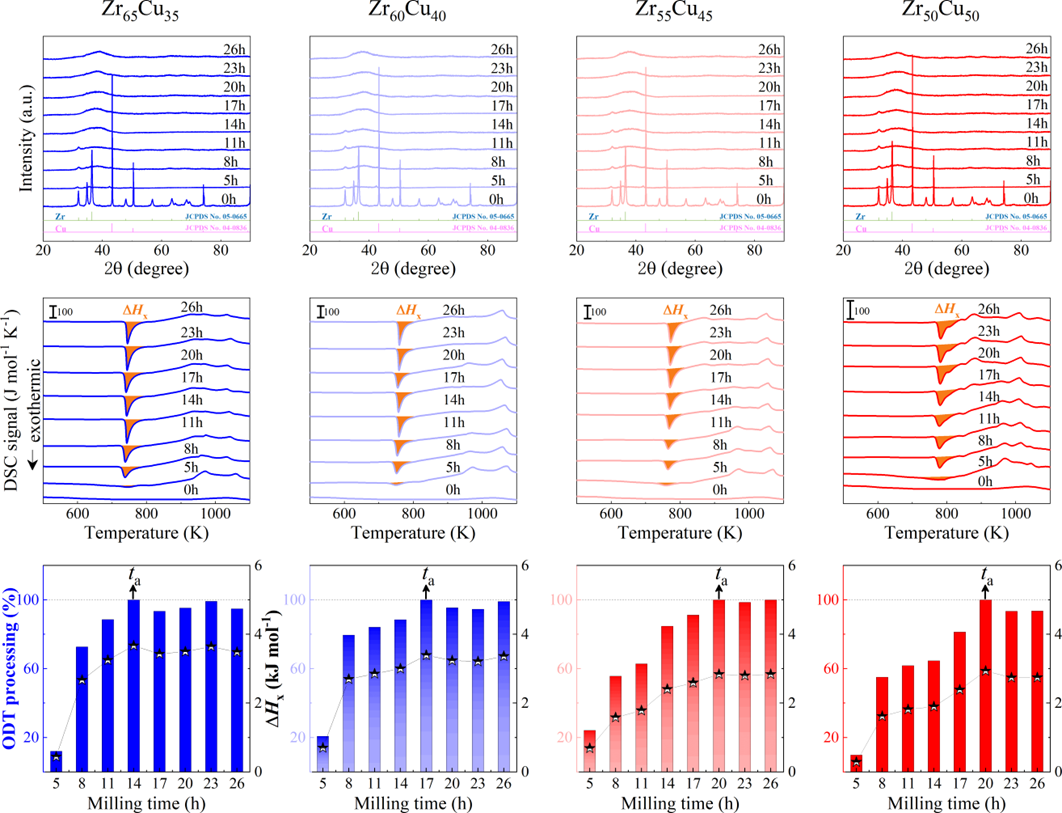}
    \caption{
    XRD pattens (upper part), DSC traces (middle part) at a heating rate of 20 K min$^{-1}$,  and processing (lower part) for mechanical
amorphization of the B-Zr$_{65}$, B-Zr$_{60}$, B-Zr$_{65}$, and B-Zr$_{50}$ alloys via high-energy ball milling of pure element powders.
    }
    \label{fig:S5}
\end{figure}

\begin{figure}[!]
    \centering
    \includegraphics[width=0.7\textwidth]{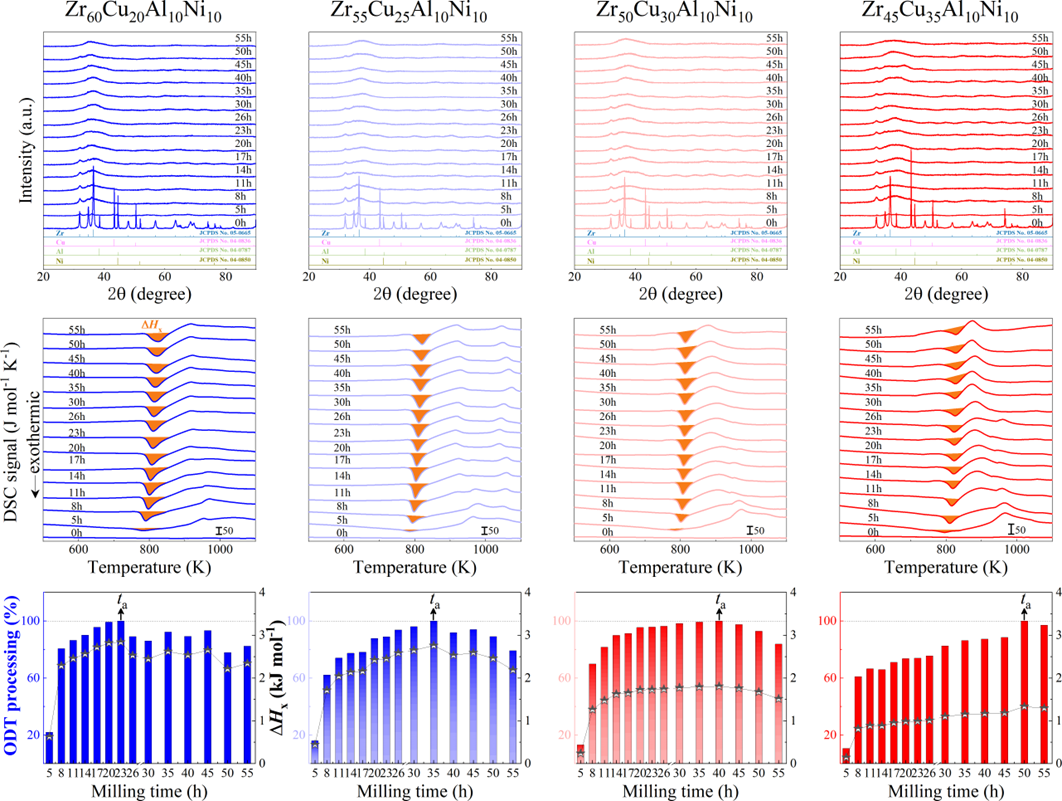}
    \caption{
     XRD pattens (upper part), DSC traces (middle part) at a heating
rate of 20 K min$^{-1}$, and processing (lower part) for mechanical
amorphization of the Q-Zr$_{60}$, Q-Zr$_{55}$, Q-Zr$_{50}$, and Q-Zr$_{45}$ alloys via high-energy ball milling of pure element powders.
    }
    \label{fig:S6}
\end{figure}

\begin{figure}[!]
    \centering
    \includegraphics[width=0.7\textwidth]{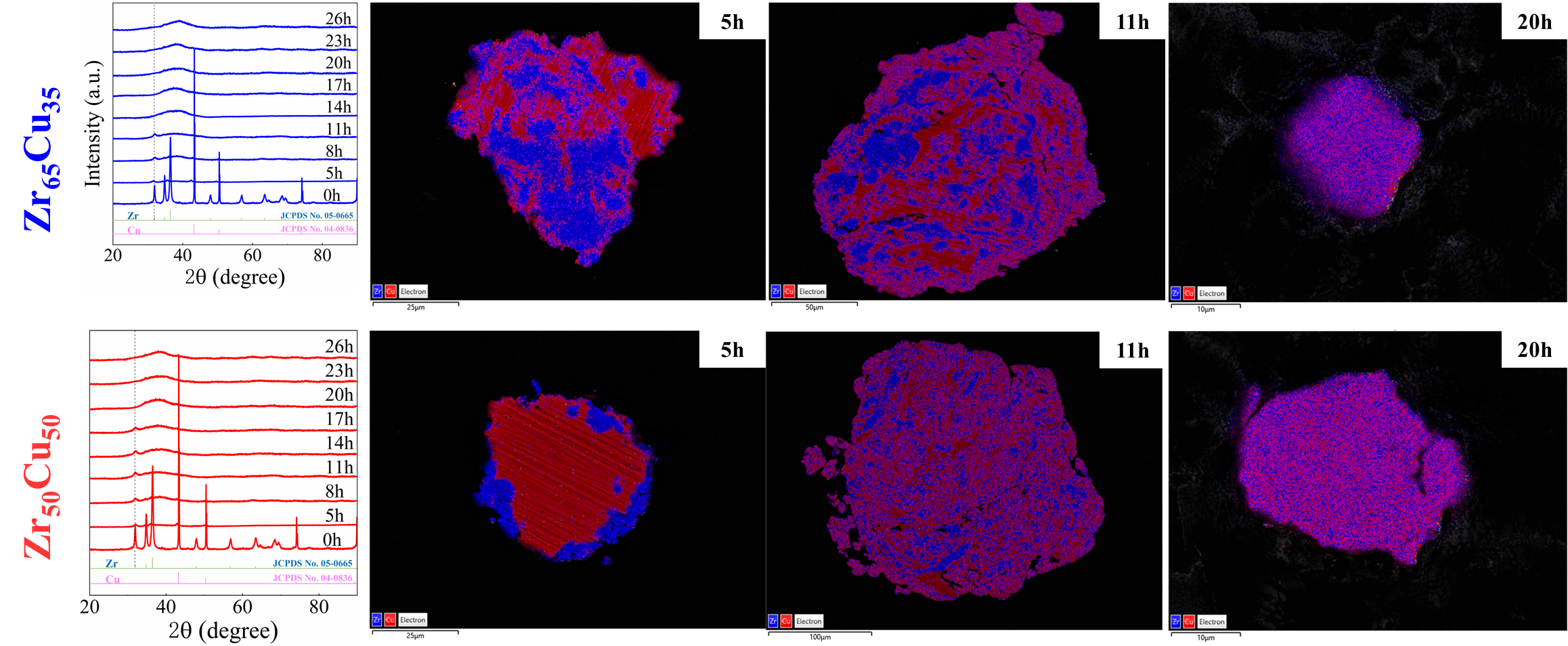}
    \caption{
    Evolution of XRD patterns and EDS mapping with the milling time for Zr$_{65}$Cu$_{35}$ (upper part) and Zr$_{50}$Cu$_{50}$ (lower part).  The distribution of Zr and Cu elements are contoured by blue and red respectively.
    }
    \label{fig:S21}
\end{figure}

In this work, the evolution of XRD patterns corresponded to a feature of type II that a decrease of elemental peaks accompanied with an increase of the amorphous broad peak for all of compositions herein (see the upper part of Figs. \ref{fig:S1}, \ref{fig:S2}, \ref{fig:S3}, \ref{fig:S4}, \ref{fig:S5}, \ref{fig:S6}). 
We supplemented the eds mapping to further detect the evolution of structures with the increase of milling time for the binary system. 
As shown in Fig. \ref{fig:S21}, each elements distributed incompatibly and no compounds formed at early milling of Zr and Cu elemental powders, corresponding to the peak intensity of pure elements decreased with the simultaneously increased amorphous halo peak. 
The identical amorphization type of here-considered alloy systems provided a prerequisite for discussion about relationship between MAA and GFA.

\subsection{Supplementary Note 6: relationship between $T_{rg}$, FWHM, and $E_c$ }

\begin{figure}[!]
    \centering
    \includegraphics[width=0.7\textwidth]{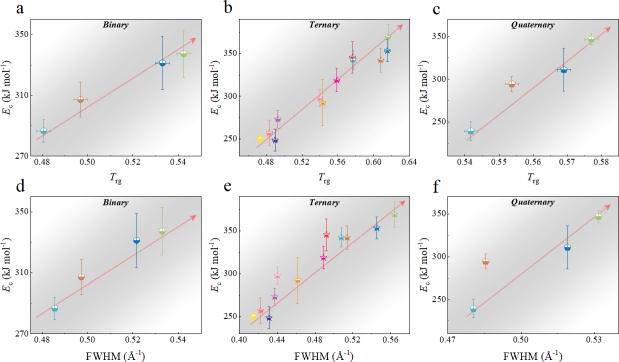}
    \caption{ Plots of GFA indicators $T_{rg}$ \textbf{(a-c)} and FWHM \textbf{(d-f)} versus $E_c$ for binary \textbf{(a, d)}, ternary \textbf{(b, e)}, and quaternary \textbf{(c, f)} systems.
    }
    \label{fig:S23}
\end{figure}

As shown in Fig. \ref{fig:S23}, $E_c$ is well correlated with GFA indicators FWHM and $T_{rg}$, which is consistent with the reported studies[6,7]. 
In terms of ball milling, a large portion of the input work $E_\text{input}$ can be dissipated in the form of heat, a small portion of which is stored in the crystalline powder material, achieving transformation from order to disorder. 

\begin{figure}[!]
    \centering
    \includegraphics[width=0.7\textwidth]{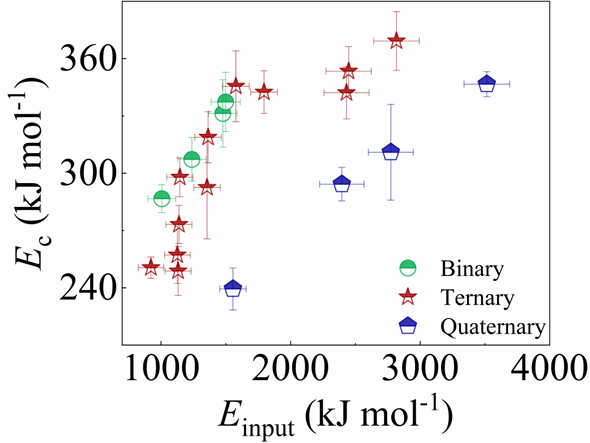}
    \caption{
     Crystallization activation energy $E_c$ versus the input work $E_\text{input}$ during ball milling for all alloys systems considered herein. 
     Error bars mean the standard deviation of data.
    }
    \label{fig:S15}
\end{figure}

To an extent, $E_\text{input}$ varied with compositions may be considered as a parameter reflecting the difficulty of amorphization transformation. That means that the poor MMA systems need more $E_\text{input}$ for amorphization during milling. Hence, the negative plot of $E_c$ with $E_\text{input}$ (Fig. \ref{fig:S15}) can be considered an energetic expression for the inverse relationship between GFA and MAA.


\pagebreak

\section{Supplementary Table}
\begin{table}[!hbpt]
    \centering
 \begin{tabular}{ccccccccc}
 \hline
Composition & $T_g$ &  $T_x$ & $T_l$ & FWHM & $T_{rg}$ & $\gamma$ & $E_c$ & $T_{P}$\\
(at. \%)  & (K) &  (K) &  (K) & (\AA$^{-1}$) & & &(kJ mol$^{-1}$) &  (K)\\
 \hline
Zr$_{65}$Cu$_{35}$  & 622  & 675  & 1294 & 0.48561  & 0.481 &  0.352  & 286.67  & 1289.8\\
Zr$_{60}$Cu$_{40}$ & 641 & 689 & 1290 & 0.49731 &  0.497 &  0.357 &  307.24 &  1277.6\\
Zr$_{55}$Cu$_{45}$ & 654 & 705 & 1227 & 0.52161 & 0.533 & 0.375 & 331.31 & 1221.6\\
Zr$_{50}$Cu$_{50}$ & 666 & 723 & 1228 & 0.53283 & 0.542 & 0.382 & 337.33 & 1223\\
 \hline
 \hline
Zr$_{72.5}$Cu$_{17.5}$Al$_{10}$ & 624 & 658 & 1318 & 0.41504 & 0.473 & 0.339 & 250.56 & 1256\\
Zr$_{70}$Cu$_{20}$Al$_{10}$ & 633 & 671 & 1309 & 0.42253 & 0.484 & 0.346 & 257.17 & 1242.9\\
Zr$_{67.5}$Cu$_{22.5}$Al$_{10}$ & 640 & 703 & 1306 & 0.43148 & 0.490 & 0.361 & 248.81 & 1204.4\\
Zr$_{65}$Cu$_{25}$Al$_{10}$ & 644 & 729 & 1306 & 0.43748 & 0.493 & 0.374 & 273.17 & 1208.1\\
Zr$_{62.5}$Cu$_{27.5}$Al$_{10}$ & 664 & 736 & 1230 & 0.44032 & 0.540 & 0.389 & 297.87 & 1223\\
Zr$_{60}$Cu$_{30}$Al$_{10}$ & 664 & 739 & 1223 & 0.46186 & 0.543 & 0.392 & 292.56 & 1211.6\\
Zr$_{57.5}$Cu$_{32.5}$Al$_{10}$ & 682 & 743 & 1220 & 0.48898 & 0.559 & 0.391 & 318.84 & 1199.3\\
Zr$_{55}$Cu$_{35}$Al$_{10}$ & 698 & 750 & 1211 & 0.49216 & 0.576 & 0.393 & 345.52 & 1190.5\\
Zr$_{52.5}$Cu$_{37.5}$Al$_{10}$ & 692 & 751 & 1199 & 0.50786 & 0.577 & 0.397 & 342.48 & 1178.9\\
Zr$_{50}$Cu$_{40}$Al$_{10}$ & 707 & 753 & 1163 & 0.51413 & 0.608 & 0.403 & 342.16 & 1143\\
Zr$_{47.5}$Cu$_{42.5}$Al$_{10}$ & 710 & 757 & 1154 & 0.54548 & 0.615 & 0.406 & 353.41 & 1149.2\\
Zr$_{45}$Cu$_{45}$Al$_{10}$ & 714 & 765 & 1159 & 0.56429 & 0.616 & 0.408 & 369.19 & 1154.8\\
 \hline
 \hline
Zr$_{60}$Cu$_{20}$Al$_{10}$Ni$_{10}$ & 662 & 757 & 1222 & 0.48048 & 0.542 & 0.402 & 239.44 & 1113.3\\
Zr$_{55}$Cu$_{25}$Al$_{10}$Ni$_{10}$ & 680 & 769 & 1228 & 0.48553 & 0.554 & 0.403 & 294.28 & 1152.4\\
Zr$_{50}$Cu$_{30}$Al$_{10}$Ni$_{10}$ & 701 & 773 & 1232 & 0.51909 & 0.569 & 0.400 & 310.99 & 1175.3\\
Zr$_{45}$Cu$_{35}$Al$_{10}$Ni$_{10}$ & 715 & 787 & 1239 & 0.53175 & 0.577 & 0.403 & 346.66 & 1190.7\\
 \hline
\end{tabular}
    \caption{ The $T_g$, $T_x$, $T_l$, FWHM, and the calculated $T_{rg}$, $\gamma$
and $E_c$ for the here-considered metallic glasses.
}
    \label{tab:S1}
\end{table}
\begin{table}[!hbpt]
    \centering
  \begin{tabular}{ccccc}
  \hline
Composition & $a\times 10^3$ & $b\times 10^6$ & $c\times 10^3$ & $d\times 10^{-6}$ \\
(at. \%) & (J g atom$^{-1}$ K$^{-2}$) & (J g atom$^{-1}$ K$^{-3}$) & (J g atom$^{-1}$ K$^{-2}$) & (J g atom$^{-1}$ K)\\
 \hline
 \hline
Zr$_{65}$Cu$_{35}$ & 2.37 & 2.00 & 4.33 & 4.46\\
Zr$_{60}$Cu$_{40}$ & 1.37 & 6.37 & 8.35 & 3.99\\
Zr$_{55}$Cu$_{45}$ & -10.05 & 13.09 & 5.33 & 4.28\\
Zr$_{50}$Cu$_{50}$ & 7.39 & 3.93 & 11.13 & 4.84\\
 \hline
 \hline
Zr$_{72.5}$Cu$_{17.5}$Al$_{10}$ & 9.14 & 0.34 & 9.22 & 5.74\\
Zr$_{70}$Cu$_{20}$Al$_{10}$ & -9.72 & 17.58 & 11.65 & 3.84\\
Zr$_{67.5}$Cu$_{22.5}$Al$_{10}$ & -5.54 & 10.66 & 10.72 & 5.33\\
Zr$_{65}$Cu$_{25}$Al$_{10}$ & -4.44 & 12.23 & 11.78 & 5.56\\
Zr$_{62.5}$Cu$_{27.5}$Al$_{10}$ & -7.77 & 16.04 & 11.87 & 5.36\\
Zr$_{60}$Cu$_{30}$Al$_{10}$ & -13.65 & 19.70 & 10.76 & 4.69\\
Zr$_{57.5}$Cu$_{32.5}$Al$_{10}$ & 0.75 & 8.15 & 10.76 & 5.43\\
Zr$_{55}$Cu$_{35}$Al$_{10}$ & 9.25 & 1.25 & 9.88 & 5.49\\
Zr$_{52.5}$Cu$_{37.5}$Al$_{10}$ & 4.09 & 4.66 & 10.09 & 5.71\\
Zr$_{50}$Cu$_{40}$Al$_{10}$ & 5.25 & 6.83 & 15.23 & 5.45\\
Zr$_{47.5}$Cu$_{42.5}$Al$_{10}$ & -3.09 & 11.27 & 14.31 & 4.40\\
Zr$_{45}$Cu$_{45}$Al$_{10}$ & 4.00 & 1.31 & 10.03 & 5.45\\
 \hline
 \hline
Zr$_{60}$Cu$_{20}$Al$_{10}$Ni$_{10}$ & 3.05 & 4.03 & 8.27 & 5.24\\
Zr$_{55}$Cu$_{25}$Al$_{10}$Ni$_{10}$ & -1.85 & 11.59 & 12.11 & 5.82\\
Zr$_{50}$Cu$_{30}$Al$_{10}$Ni$_{10}$ & -0.09 & 10.65 & 13.00 & 5.82\\
Zr$_{45}$Cu$_{35}$Al$_{10}$Ni$_{10}$ & 5.23 & 1.87 & 7.52 & 8.50\\
 \hline
\end{tabular}
    \caption{ Fitting constants for the heat capacity data, using
$C_P^c (T) = 3R + aT + bT^2$ to fit the crystalline state heat capacity data and
$C_P^l (T) = 3R + cT + dT^{-2}$ to fit the liquid heat capacity data.
}
    \label{tab:s2}
\end{table}

\pagebreak

\section{ Supplementary Figures}

 \begin{figure}[!]
     \centering
     \includegraphics[width=0.7\textwidth]{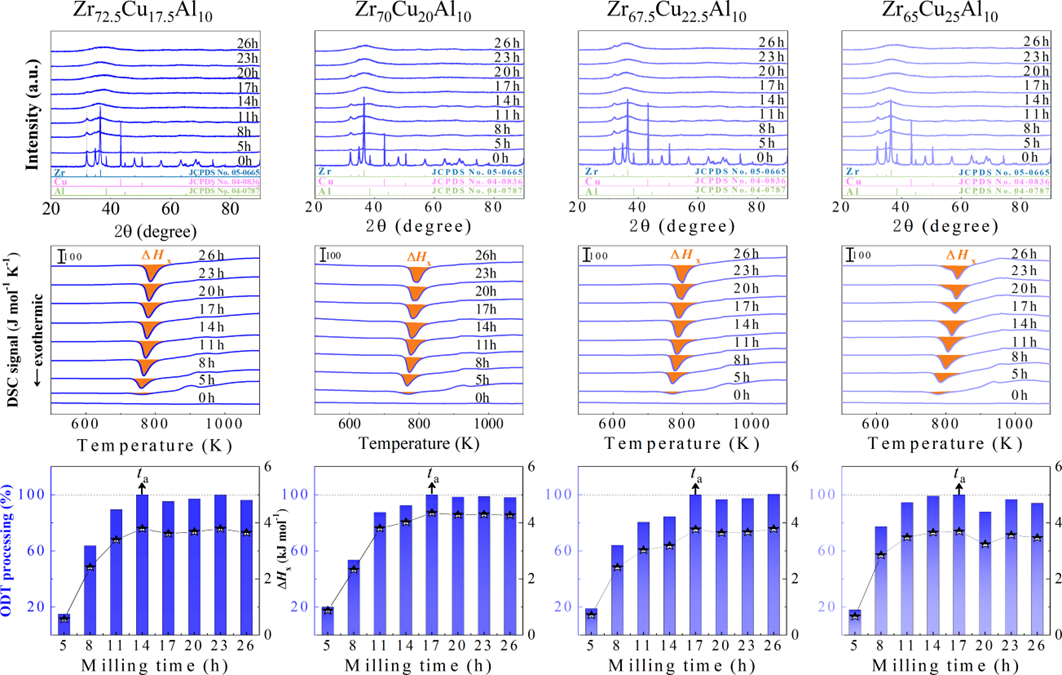}
     \caption{ XRD pattens (upper part), DSC traces (middle part) at a heating rate of 20 K min$^{-1}$, and processing (lower part) for mechanical amorphization of the T-Zr$_{72.5}$, T-Zr$_{70}$, T-Zr$_{67.5}$, and T-Zr$_{65}$ alloys via high-energy ball milling of pure element powders.
     }
     \label{fig:S1}
 \end{figure}
\begin{figure}[!]
    \centering
    \includegraphics[width=0.7\textwidth]{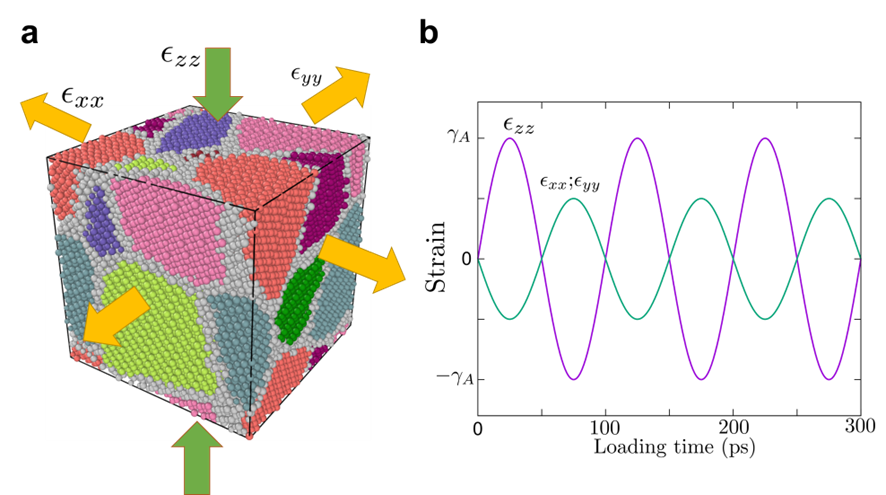}
    \caption{
    The schematic of the loading process. 
    \textbf{a} Illustration of the loading direction for the CuZr polycrystal system.  Different grains and grain boundaries are distinguished by various colors. 
\textbf{b} Three strains are shown as functions of loading time for $T_p = 100$ ps and amplitude strain $\gamma_A$ .
}
    \label{fig:S7}
\end{figure}
\begin{figure}[!]
    \centering
    \includegraphics[width=0.8\textwidth]{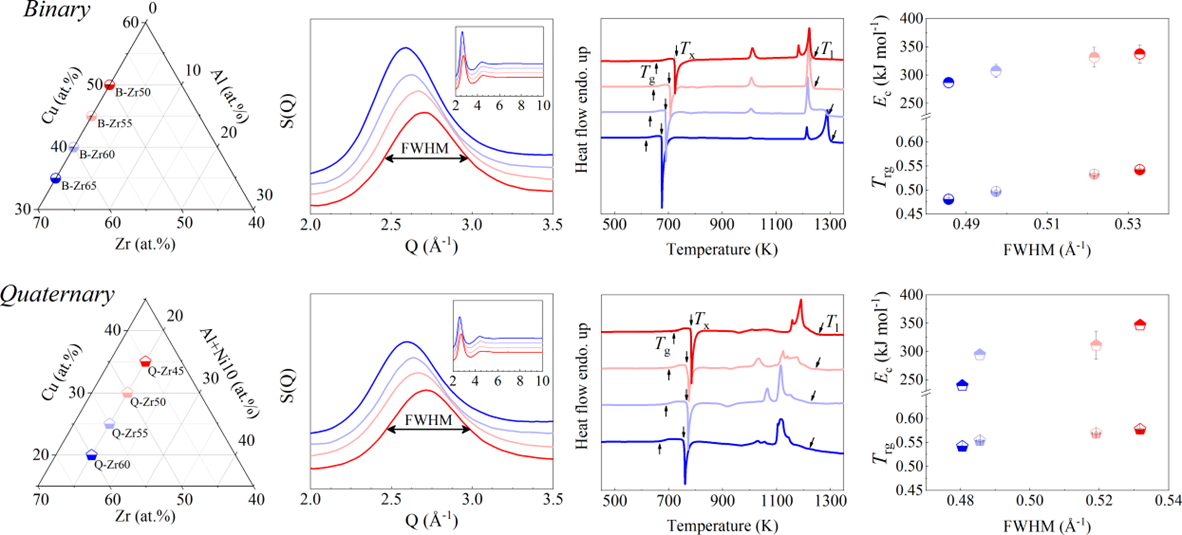}
    \caption{
    Compositions contours, S(Q) obtained from neutron diffraction, DSC traces at a heating rate of 20 K min$^{-1}$, and GFA indicators relationships ($T_{rg}$, $E_c$, and FWHM) for binary ZrCu and quaternary ZrCuAlNi alloy systems considered herein. Error bars mean the standard deviation of data.
}
    \label{fig:S9}
\end{figure}
\begin{figure}[!]
    \centering
    \includegraphics[width=0.7\textwidth]{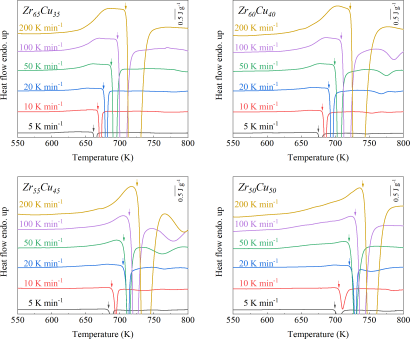}
    \caption{
    DSC traces at various heating rates of 5, 10, 20, 50, 100, and 200 K min$^{-1}$ for binary ZrCu alloy systems. Crystallization temperature dependent on heating rates is labeled by arrows.
    }
    \label{fig:S10}
\end{figure}
\begin{figure}[!]
    \centering
    \includegraphics[width=0.7\textwidth]{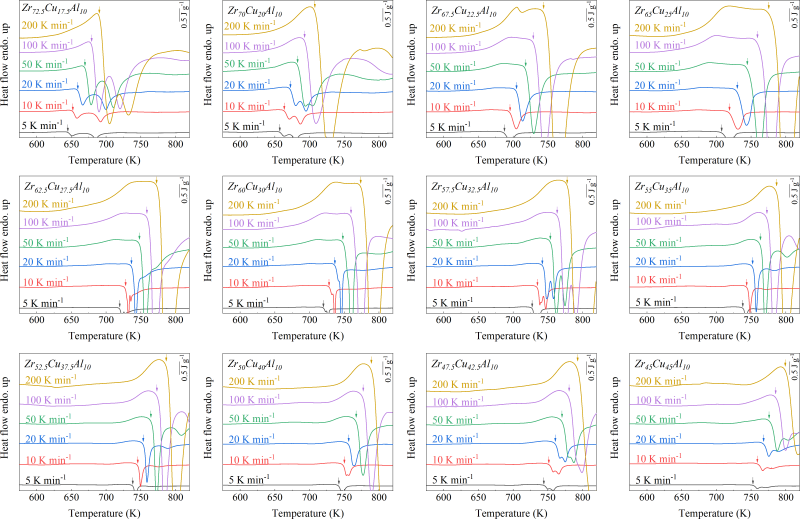}
    \caption{
     DSC traces at various heating rates of 5, 10, 20, 50, 100, and 200 K min$^{-1}$ for ternary ZrCuAl alloy systems. Crystallization temperature dependent on heating rates is labeled by arrows.
    }
    \label{fig:S11}
\end{figure}
\begin{figure}[!]
    \centering
    \includegraphics[width=0.7\textwidth]{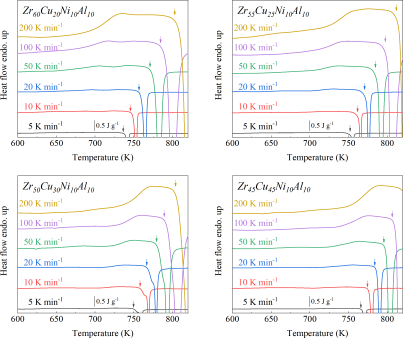}
    \caption{
    DSC traces at various heating rates of 5, 10, 20, 50, 100, and 200 K min$^{-1}$ for quaternary ZrCuNiAl alloy systems. Crystallization temperature dependent on heating rates is labeled by arrows.
   }
    \label{fig:S12}
\end{figure}
\begin{figure}[!]
    \centering
    \includegraphics[width=0.7\textwidth]{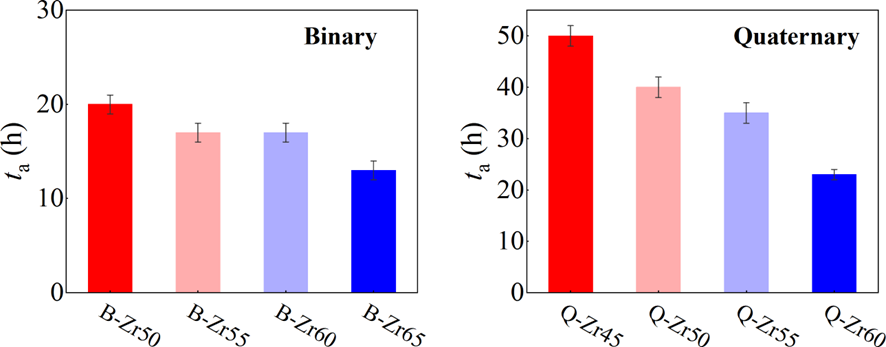}
    \caption{
     MAA indicator (amorphization time, $t_a$) for binary ZrCu and quaternary ZrCuAlNi alloy systems considered herein. Error bars mean the standard deviation of data.
    }
    \label{fig:S14}
\end{figure}
\begin{figure}[!]
    \centering
    \includegraphics[width=0.7\textwidth]{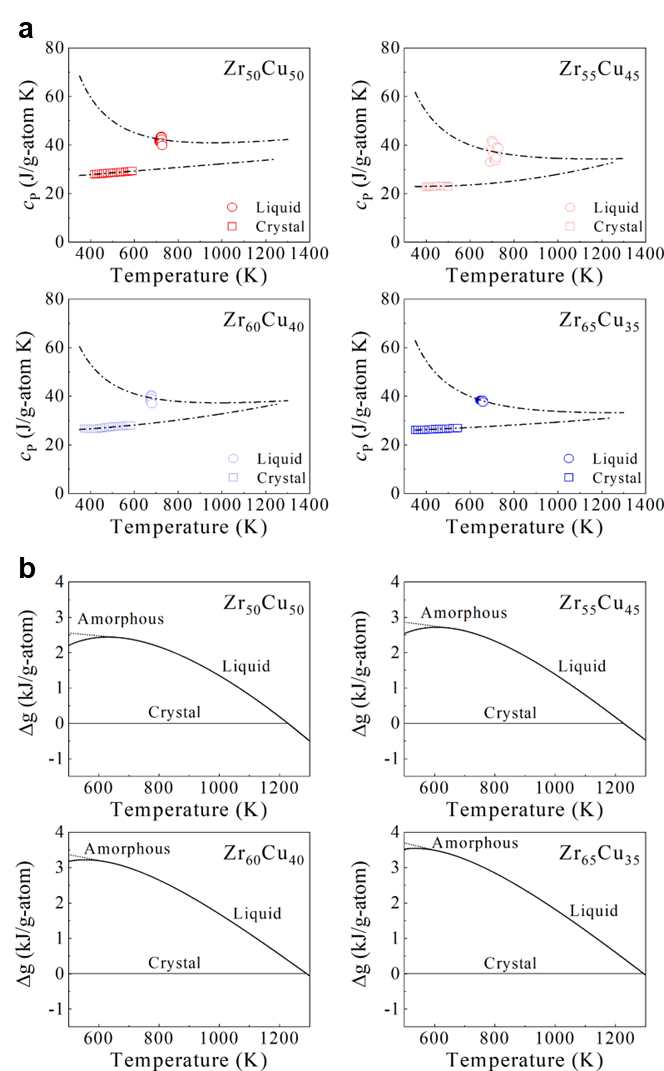}
    \caption{
 \textbf{a} Specific heat capacity of the undercooling liquid (circle) and the crystal (square) for binary alloy systems. The dot dash lines represent the fits using Eqs. (4) and (5), respectively. \textbf{b} Gibbs free energy of the undercooled liquid with respect to the crystal as a function of temperature for binary alloy systems.
    }
    \label{fig:S16}
\end{figure}
\begin{figure}[!]
    \centering
    \includegraphics[width=0.7\textwidth]{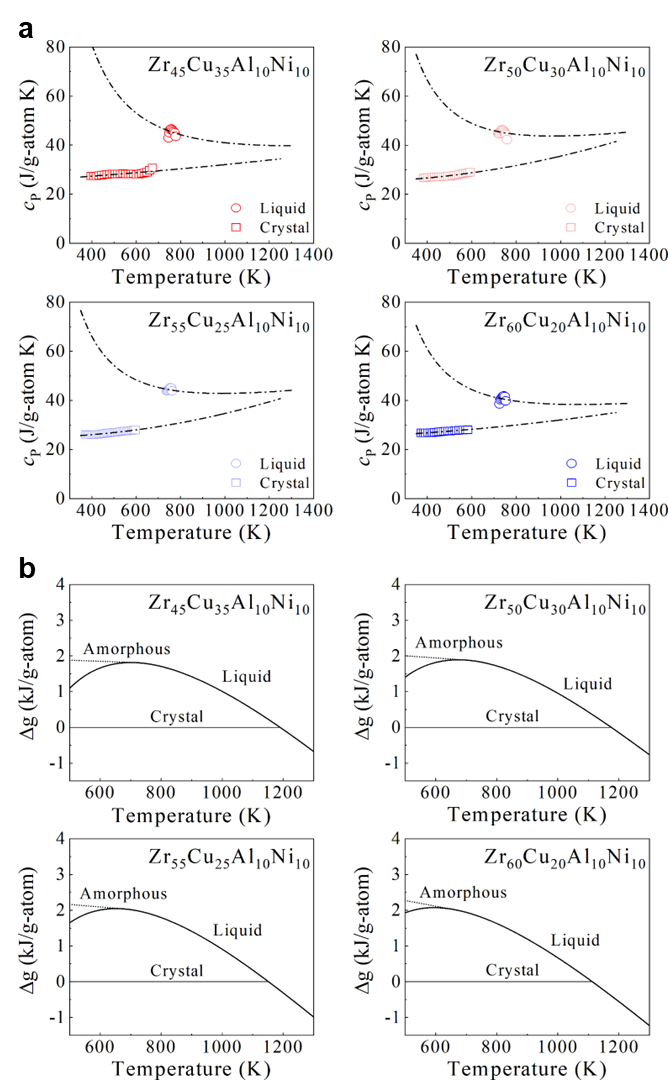}
    \caption{
    \textbf{a} Specific heat capacity of the undercooling liquid (circle) and the crystal (square) for quaternary alloy systems. The dot dash lines represent the fits using Eqs. (4) and (5), respectively.
    \textbf{b} Gibbs free energy of the undercooled liquid with respect to the crystal as a function of temperature for quaternary alloy systems.
    }
    \label{fig:S17}
\end{figure}
\begin{figure}[!]
    \centering
    \includegraphics[width=0.7\textwidth]{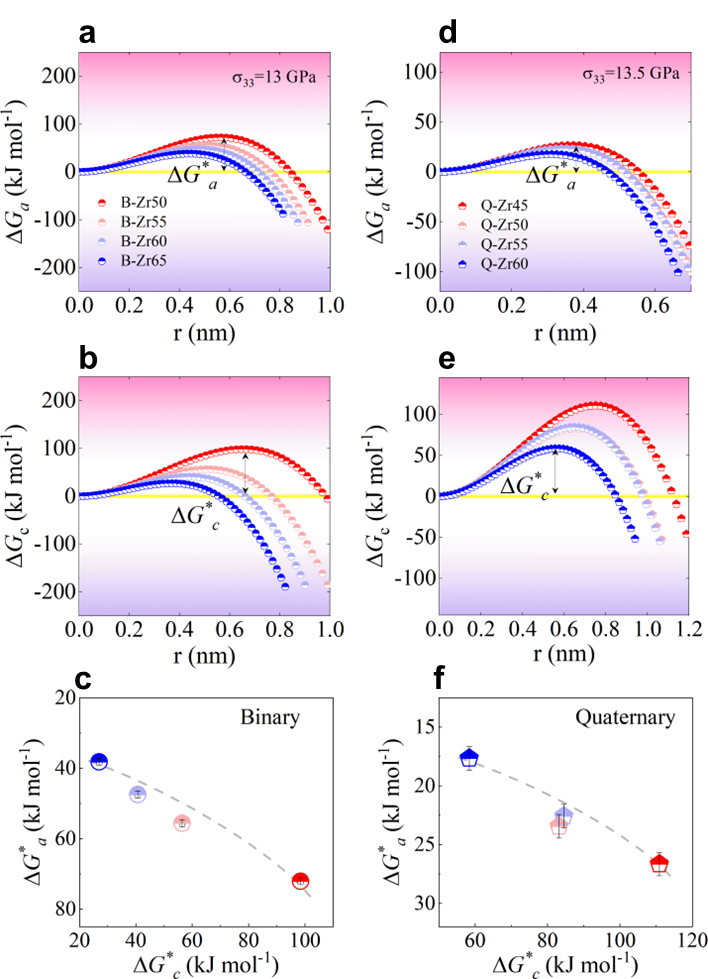}
    \caption{
     Gibbs free-energy change associated with nucleation of an \textbf{a, d}
amorphous and \textbf{b, e} crystal embryo for \textbf{a, b} binary and \textbf{d, e} quaternary alloy systems, respectively. \textbf{c, f} Correlation of $\Delta G_a^*$ and $\Delta G_c^*$ for \textbf{c} binary and \textbf{f} quaternary alloy systems.
    }
    \label{fig:S18}
\end{figure}
\begin{figure}[!]
    \centering
    \includegraphics[width=0.7\textwidth]{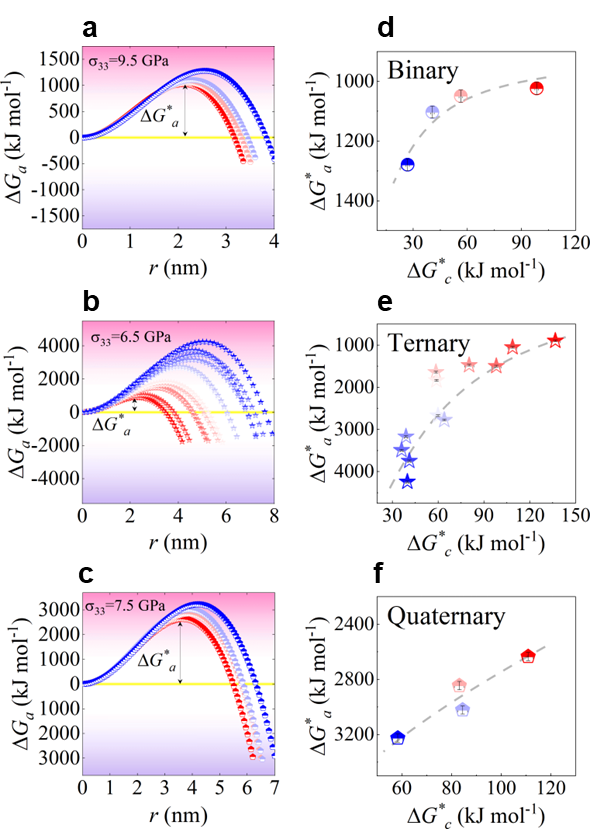}
    \caption{
     \textbf{(a)-(c)} Gibbs free-energy change associated with nucleation of an amorphous embryo and  \textbf{(d)-(f)}  correlation of $\Delta G_a^*$ and $\Delta G_c^*$ for \textbf{(a), (d)} binary, \textbf{(b), (e)} ternary,  and \textbf{(c), (f)} 
     quaternary alloy systems at low external stress, respectively.
     }
    \label{fig:S19}
\end{figure}
\begin{figure}[!]
    \centering
    \includegraphics[width=0.7\textwidth]{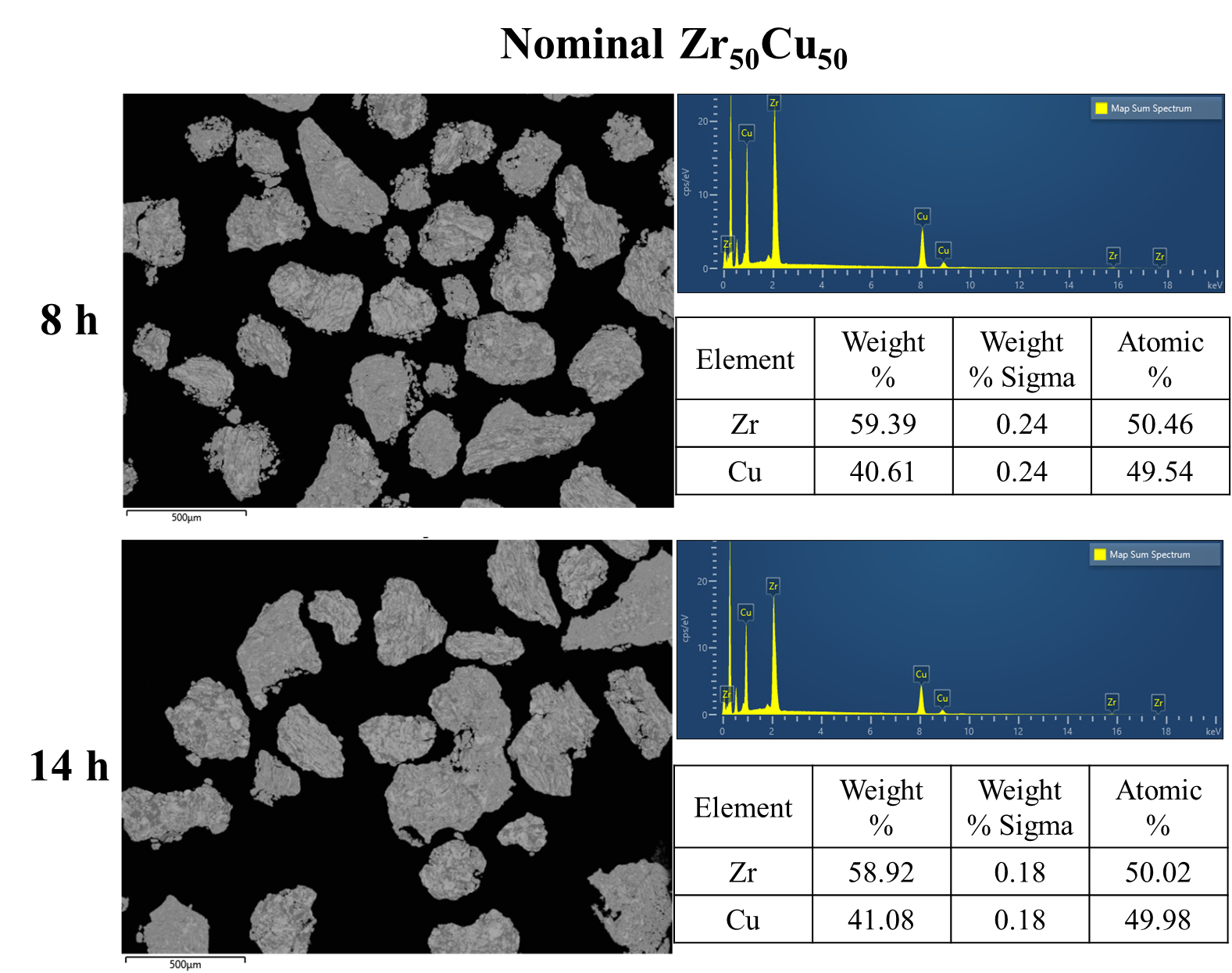}
    \caption{
    SEM images and the element spectrum and the related atomic
ratio at a milling time of 8 and 14 h for Zr$_{50}$ Cu$_{50}$ systems. The light and grey regions of SEM images represent Zr and Cu respectively.
    }
    \label{fig:S22}
\end{figure}
\begin{figure}[!]
    \centering
    \includegraphics[width=0.7\textwidth]{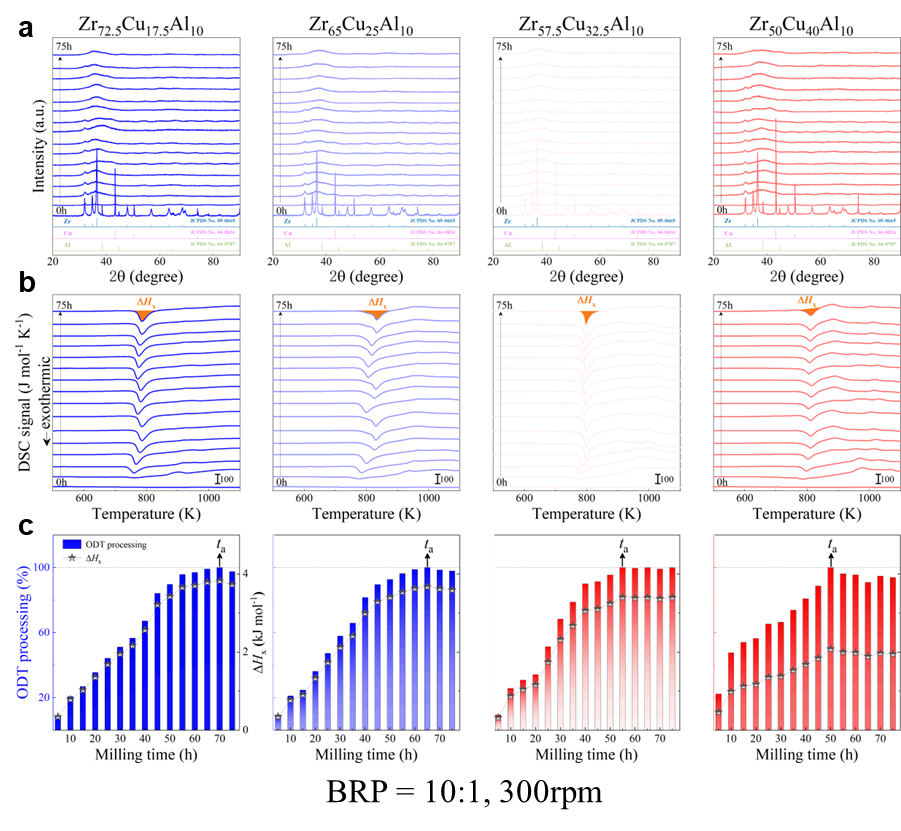}
    \caption{\textbf{(a)} XRD pattens, \textbf{(b)} DSC traces at a heating rate of 20 K min$-1$,and \textbf{(c)} processing for mechanical amorphization of the T-Zr$_{72.5}$, T-Zr$_{65}$,T-Zr$_{57.5}$, and T-Zr$_{50}$ alloys with a milling condition of BRP 10:1 and 300 rpm.
    }
    \label{fig:S24}
\end{figure}
\begin{figure}[!]
    \centering
    \includegraphics[width=0.7\textwidth]{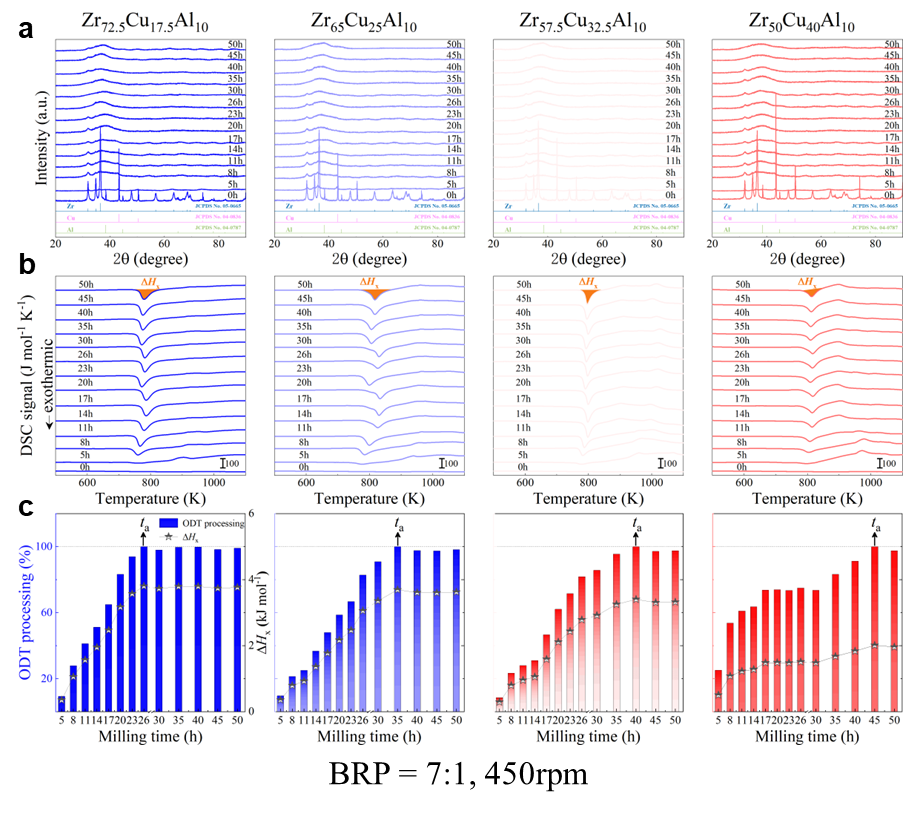}
    \caption{
     \textbf{(a)} XRD pattens, \textbf{(b)} DSC traces at a heating rate of 20 K min $^{-1}$, and (c) processing for mechanical amorphization of the T-Zr$_{72.5}$, T-Zr$_{65}$, T-Zr$_{57.5}$, and T-Zr$_{50}$ alloys with a milling condition of BRP 7:1 and 450 rpm.
    }
    \label{fig:S25}
\end{figure}
\begin{figure}[!]
    \centering
    \includegraphics[width=0.7\textwidth]{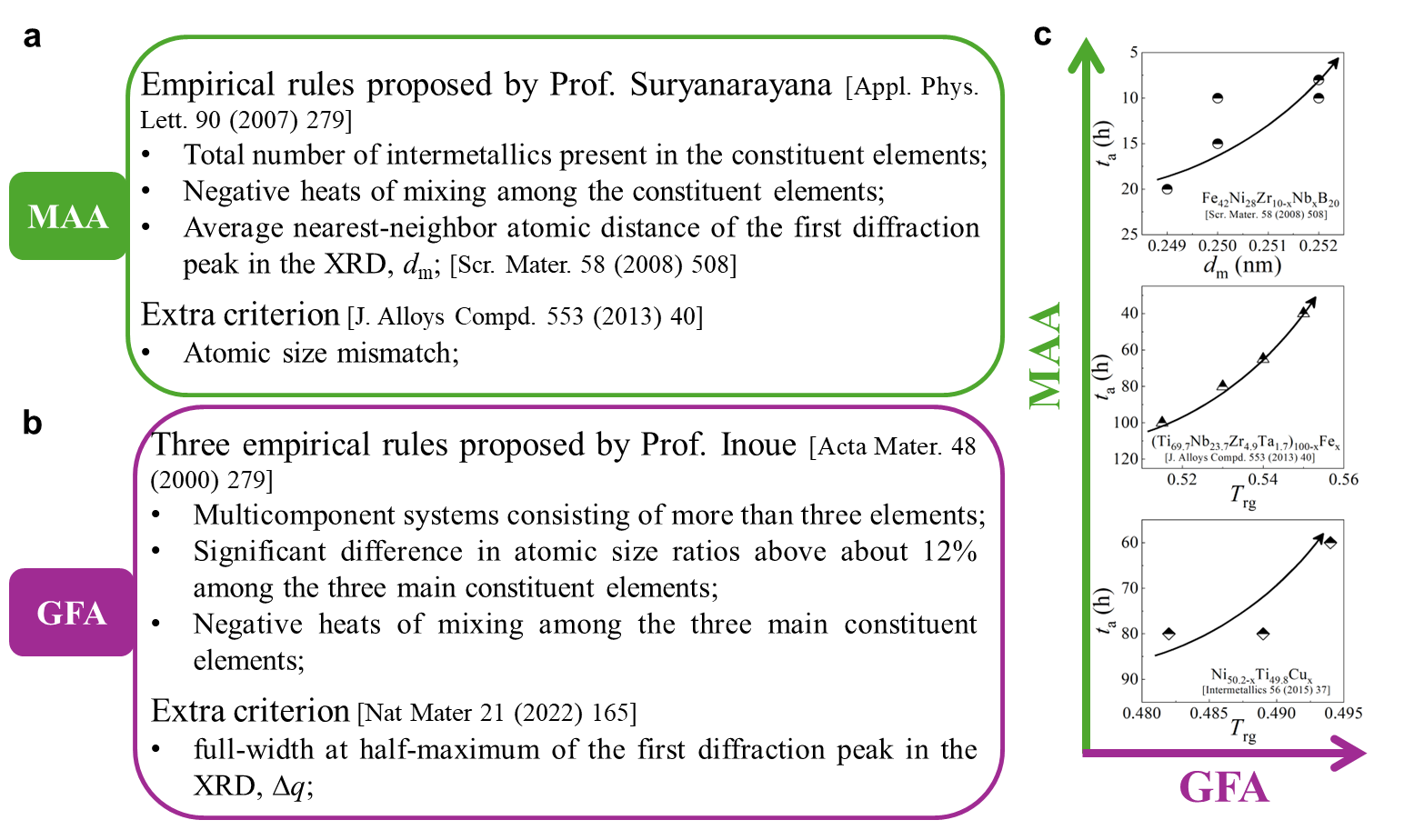}
    \caption{ The similar empirical criteria for MAA \textbf{(a)} and GFA \textbf{(b)}. 
    \textbf{(c)} The previous experimental data showed a positive relationship between MAA and GFA.
    }
    \label{fig:S26}
\end{figure}
\begin{figure}[!]
    \centering
    \includegraphics[width=0.7\textwidth]{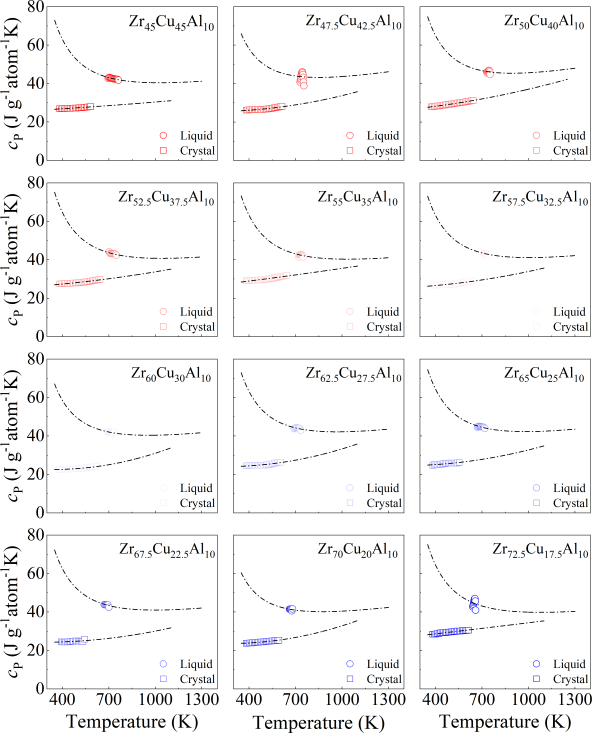}
    \caption{
    Specific heat capacity of the undercooling liquid (circle) and the crystal (square) for ternary alloy systems. The dot dash lines represent the fits using Eqs. (5) and (6), respectively.
    }
    \label{fig:S27}
\end{figure}
\begin{figure}[!]
    \centering
    \includegraphics[width=0.7\textwidth]{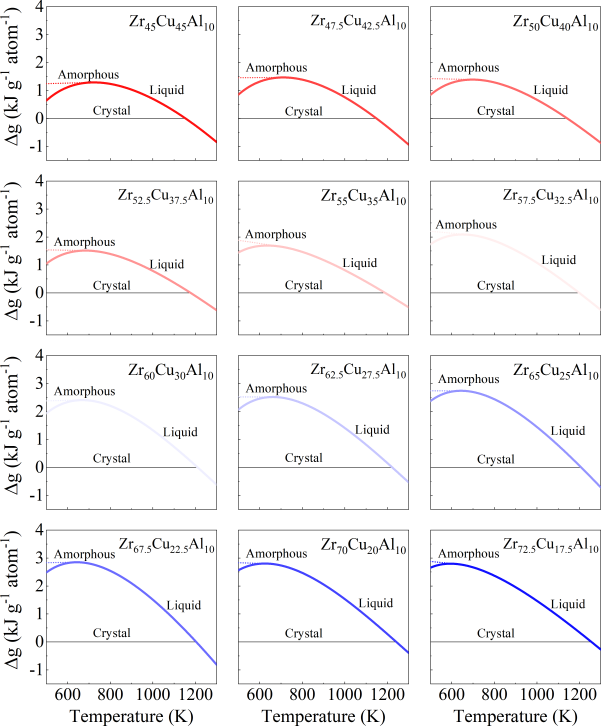}
    \caption{Gibbs free energy of the undercooled liquid with respect to the
crystal as a function of temperature for ternary alloy systems.
}
    \label{fig:S28}
\end{figure}
\begin{figure}[!]
    \centering
    \includegraphics[width=0.5\textwidth]{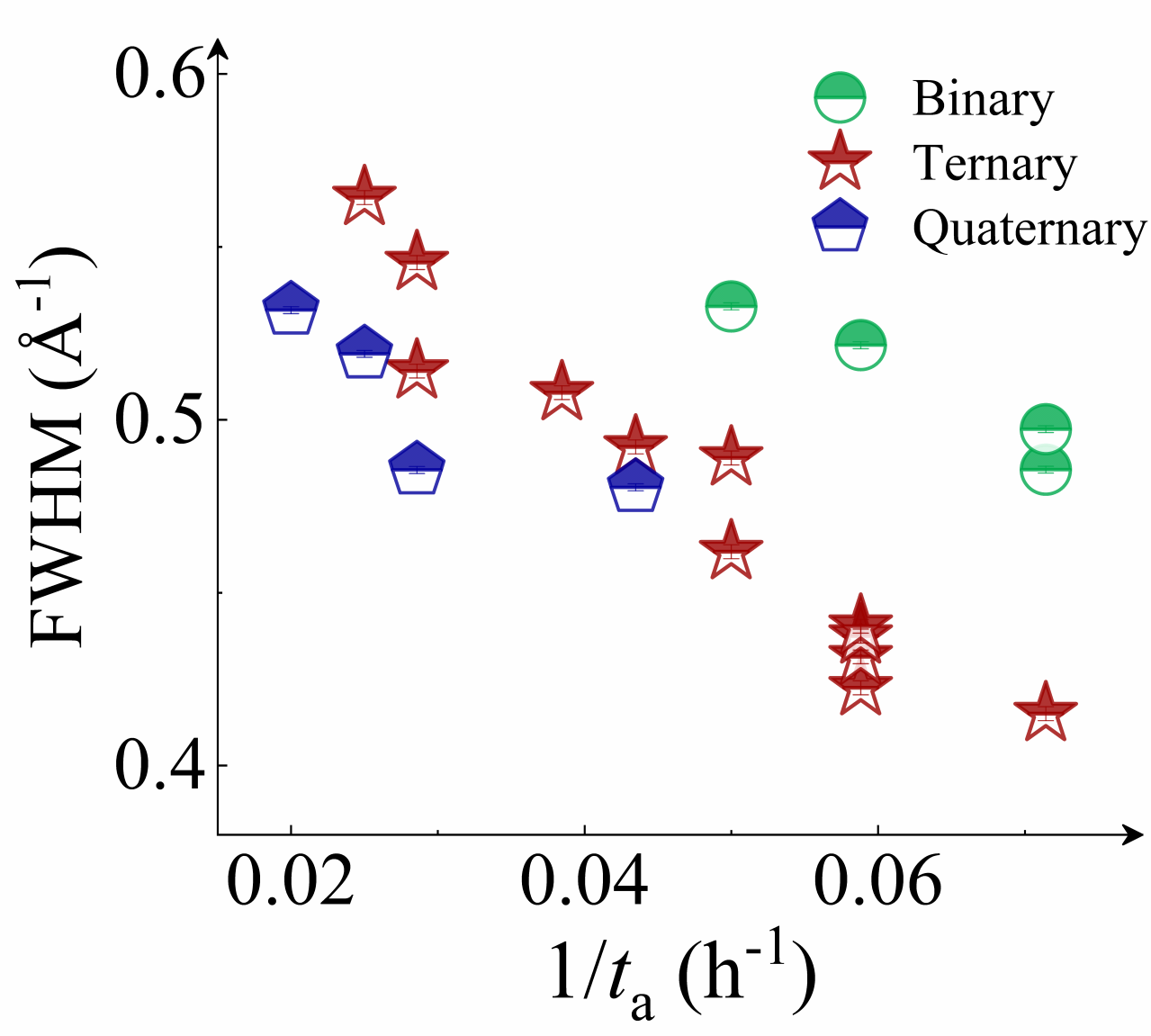}
    \caption{Anomalous correlations exist in plots of FWHM versus $1/t_a$ for
all alloy systems considered here.
}
    \label{fig:S29}
\end{figure}

\end{document}